\documentclass[twocolumn,english]{revtex4-2}
\usepackage[utf8]{inputenc}
\usepackage{xcolor}
\usepackage{float}
\usepackage{textcomp}
\usepackage{amsmath}
\usepackage{amssymb}
\usepackage{stmaryrd}
\usepackage{graphicx}
\PassOptionsToPackage{normalem}{ulem}
\usepackage{hyperref}
\hypersetup{colorlinks=true, linkcolor=blue, citecolor=magenta, urlcolor=blue}

\usepackage{ulem}

\makeatletter

\@ifundefined{showcaptionsetup}{}{%
 \PassOptionsToPackage{caption=false}{subfig}}
\usepackage{subfig}
\makeatother

\usepackage{babel}
\begin{document}
\global\long\def\sgn{\mathrm{sgn}}%
\global\long\def\ket#1{\left|#1\right\rangle }%
\global\long\def\bra#1{\left\langle #1\right|}%
\global\long\def\sp#1#2{\langle#1|#2\rangle}%
\global\long\def\abs#1{\left|#1\right|}%
\global\long\def\avg#1{\langle#1\rangle}%

\title{Effect of electron-electron interactions on the propagation of ultrashort voltage pulses in a Mach-Zehnder interferometer}

\author{Prasoon Kumar$^{1}$}
\author{Thomas Kloss$^{2}$}
\author{Xavier Waintal$^1$}
\email{xavier.waintal@cea.fr}
\address{$^1$Universit\'e Grenoble Alpes, CEA, Grenoble INP, IRIG, PHELIQS, 38000 Grenoble, France}
\address{$^2$Universit\'e Grenoble Alpes, CNRS Grenoble, Grenoble INP, Institut N\'eel, 38000 Grenoble, France}

\date{\today}
\begin{abstract}
Electronic interferometers have been identified as possible candidates for building electronic flying qubits. Such a regime requires ultrafast voltage pulses
whose duration is shorter than the time of flight through the device. 
Understanding the corresponding physics in the presence of such short excitations requires  a proper treatment of electron-electron interactions.
In this article, we take a step in this direction by performing time-resolved
simulations of a Mach-Zehnder interferometer treating the interactions
at the time-dependent mean-field level. We find that the main effect of the interaction is the renormalization of the pulse velocity. Very importantly, the interference effects appear to be robust to the presence of interactions.
\end{abstract}
\maketitle

%%%%%%%%%%%%%%%%%%%%%%%%%%%%%%%%%%%%%%%%%%%%%%%%%%%%%%%%%%%%%%%%%%%%%%%%%%%%%
\section{Introduction}
%%%%%%%%%%%%%%%%%%%%%%%%%%%%%%%%%%%%%%%%%%%%%%%%%%%%%%%%%%%%%%%%%%%%%%%%%%%%%

\begin{figure}[H]
\includegraphics[width= \linewidth]{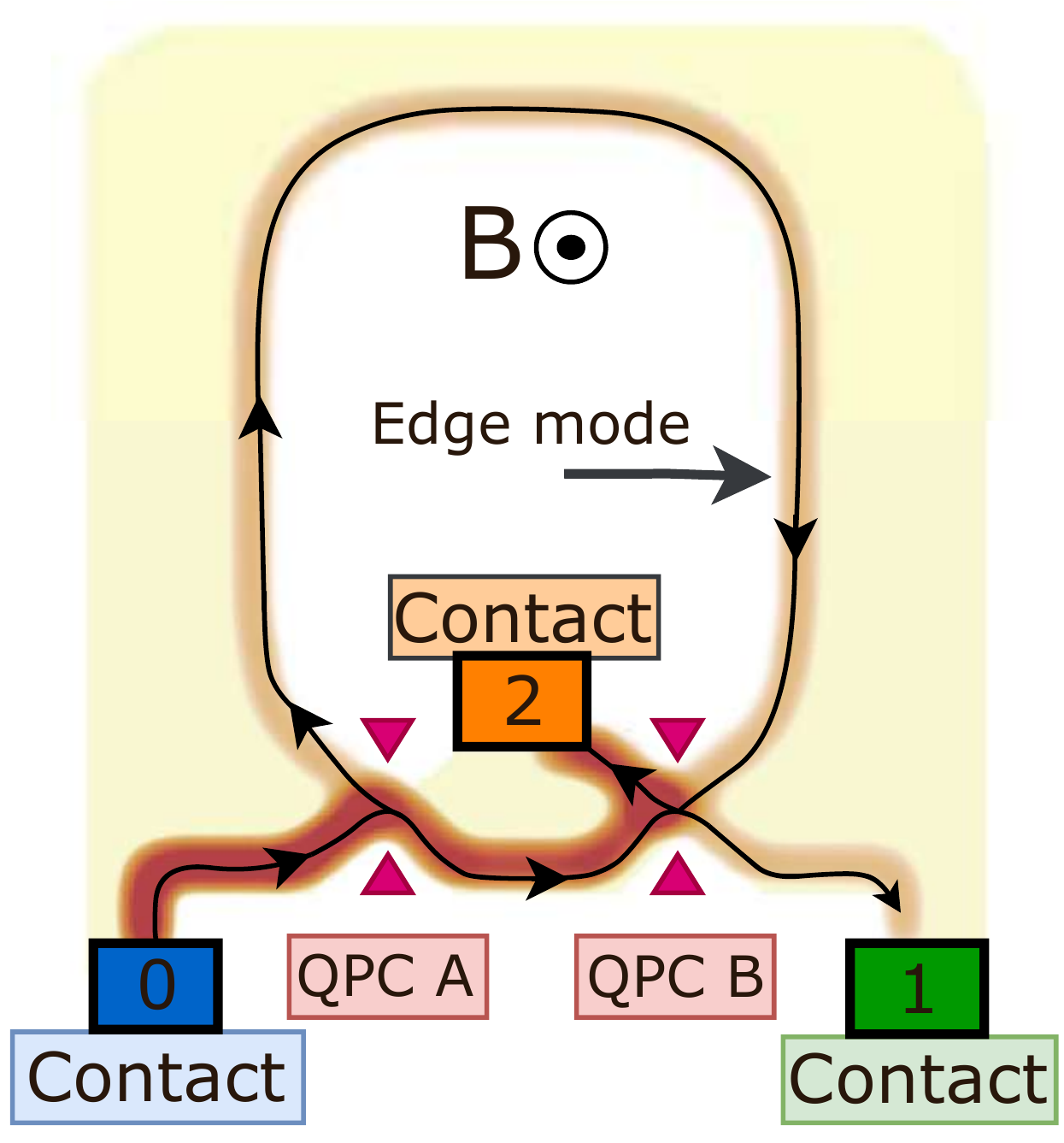}
\caption{\label{fig:system}Schematic of a Mach-Zehnder interferometer in the quantum Hall regime. The device consists of a ring shaped 2DEG connected to three contacts: two outer contacts (0 and 1) and one inner contact (2). Quantum point contacts (QPC A and QPC B) act as beam splitters for the chiral edge state injected from source contact 0. The edge channel is partially transmitted and reflected at QPC A, with the two paths recombining at QPC B. The resulting current is collected at the inner contact 2 and outer contact 1.}
\end{figure}

One key frontier in phase-coherent nanoelectronics is what happens at ultrahigh frequency, in the terahertz regime. Indeed, the typical phase-coherent length $L_\phi$ in a two dimensional electron gas (2DEG) is of the order of a few micrometers \cite{Roulleau2008} while a charged excitation propagates at a speed $v_P \sim 10^5 m/s$
or faster \cite{Roussely2018}. Hence, the duration of a voltage pulse must be shorter than $L_\phi/v_P\sim 10$ ps for dynamical electronic interference effects to be able to take place. It is a very active field of research to try and reach this regime experimentally; it is also a necessary step towards building an electronic flying qubit \cite{Bauerle2018}.

When and if such a regime can be reached, some rather counter-intuitive phenomena could be engineered including \cite{Levitov1996,Levitov1997,Levitov2006,Dubois2013}. 
Among those, an interesting alley is that of controlling an interference pattern dynamically using voltage pulses \cite{Gaury2014a,Gaury2014b,Gaury2015,Portugal2024,Saha2025,Kloss2025}. The predicted effects include DC currents that \emph{oscillate} with the amplitude of the voltage bias \cite{Gaury2014b}, the non-superconducting analogue to the AC-Josephson effect \cite{Gaury2015} or a current that oscillates as a function of the time waited before a pulse is released inside the device \cite{Gaury2014a}.

All these predictions, however, have been performed in non-interacting models. 
The reason is chiefly technical: treating correlations is difficult while numerical simulations must be performed for large devices and long times. 
Indeed the latter is necessary for the results to be controlled by the low energy physics and not by atomistic effects such as band gaps whose characteristic time-scale $\sim 1$ fs is much shorter. 
Yet, it is well known that the transient presence of charges inside a nanoelectronics devices must be accounted for with a treatment of electron-electron interactions, if only to account for displacement currents properly. 
This point was discussed early on by Büttiker \cite{Buttiker1993,Buttiker1995} who noted that the
AC-conductance of a non-interacting theory did not obey current conservation (the sum of incoming AC currents does not match the sum of out-going AC current) nor gauge-invariance (raising the voltage on all electrodes generates some non-zero currents). 
The breakdown of these two important laws is the manifestation that
the electrostatics  of the problem has not been treated properly.
In the non-interacting theory, charges are allowed to pile up in parts of the system at great cost in Coulomb energy. A proper treatment of
Coulomb repulsion would have them screened to keep the system globally neutral. It follows that Coulomb repulsion is expected to affect the physics quite significantly. The goal of this article is to provide such a description at the time-dependent mean-field level, the minimum level of description that can treat displacement currents properly while still remaining feasible numerically. Specifically, we study how the propagation of a voltage pulse inside a Mach-Zehnder interferometer (MZI), and the associated predicted effects, is affected by electron-electron interactions. Our chief conclusion is that the interference effects are robust to the presence of interaction, at the level treated in this work.

This article is structured as follows: Section \ref{sec:review}
presents the MZI and reviews previous non-interacting results on the propagation of pulse in this system. Section \ref{sec:model} discusses the approach used in our simulations to simulate the propagation of pulses in this MZI. Before performing simulations of the full device, we first break it into smaller units that we study separately. In Section \ref{sec:1D-Wire} and \ref{sec:QPC}, we characterize two of such units: the quasi-one dimensional wire and the quantum point contact at the DC and pulse level. The simulation for the full MZI are presented in Section \ref{sec:MZI}.

%%%%%%%%%%%%%%%%%%%%%%%%%%%%%%%%%%%%%%%%%%%%%%%%%%%%%%%%%%%%%%%%%%%%%%%%%%%%%
\section{Review of the non-interacting theory}
\label{sec:review}
%%%%%%%%%%%%%%%%%%%%%%%%%%%%%%%%%%%%%%%%%%%%%%%%%%%%%%%%%%%%%%%%%%%%%%%%%%%%%

We start by describing the problem of interest and reviewing the corresponding non-interacting predictions.
A schematic of the system under consideration is shown in Fig. \ref{fig:system}.
It consists of a 2DEG in a perpendicular magnetic field $B$ that brings it to the $\nu=1$ quantum Hall plateau.  The 2DEG has a ring topology and is connected to three Ohmic contacts: two outer contacts and one inner contact, following the design used in the experiments 
\cite{Roulleau2008,Heiblum2003}. The electronic propagation entirely takes place inside the chiral edge states; the figure shows the edge state originating from contact $0$, the counter propagating one originating from contact $1$ is not shown, nor is the one from $2$. Two quantum point contacts serve as beam splitters for the chiral edge states. A voltage pulse is sent to contact $0$ and the transmitted (resp. reflected) currents $I_1$ (resp. $I_2$) are measured in contact $1$ (resp. $2$)

In DC, the behaviour of such a MZI is well understood: the edge state originating from contact $0$ is split at the first QPC (A) into a short lower path of length $L_l$ and a long upper path of length $L_u$ that 
goes around the whole inner part of the ring. The two paths are recombined  at QPC B and continue to either contact 1 (which we call the transmitted path) or to contact $2$ (which we call the reflected path). We call $d_{\rm A}$ (resp. $D_{\rm A}=|d_{\rm A}|^2$) the transmission amplitude (resp. probability) of QPC A
and $r_{\rm A}$ (resp. $R_{\rm A}=1-D_{\rm A}$) the corresponding reflection amplitude (resp. probability) with similar notations for QPC B. The total transmission amplitude $d_{\rm tot}$ to contact $1$ (up to a global phase) is given by the interference of the two paths:
\begin{equation}
  d_{\rm tot} = d_{\rm A }d_{\rm B }+ r_{\rm A }r_{\rm B }e^{i\Phi} 
\end{equation}
where
\begin{equation}
    \Phi = k_F (L_u-L_d) + \frac{BS}{\hbar/e}
\end{equation}
and $k_F$ is the Fermi momentum in the edge state and $S$ the surface of the inner ring (more precisely the surface enclosed by the two edge states). It follows that the differential conductance given by the Landauer formula $dI_1/dV_0= e^2/h |d_{\rm tot}|^2$ is expected to present oscillations with the magnetic field or a side gate that changes the surface $S$ of the interfering loop,
\begin{align}
 D_{\rm tot} = |d_{\rm tot}|^2 = D_{\rm A } D_{\rm B } + (1-D_{\rm A })
 (1-D_{\rm B }) +\nonumber \\
 2 \cos\Phi \sqrt{D_{\rm A } D_{\rm B } (1-D_{\rm A })
 (1-D_{\rm B })}
 \label{eq:dc_mzi}
\end{align}
Such electronic interference effects have been repeatedly observed experimentally and obey well the above non-interacting theory up to a damping  factor $e^{-L/L_\phi}$ due to decoherence processes. 

Let us now consider what happens when we send a voltage pulse of duration $\tau_P$ and amplitude $V_P$ described by $V_0(t)$.
A typical form that we will consider is a Gaussian pulse,
\begin{equation}
\label{eq:gauss_pulse}
    V_0(t) = V_P e^{-t^2/\tau_P^2}
\end{equation}
but Lorentzian pulses (also known as Levitons) are also very popular.
For the purpose of this work, the precise shape is mostly irrelevant.
What matters are $\tau_P$ and $V_P$. We are interested in the total number $n_1$ of transmitted electrons,

\begin{equation}
    n_1 \equiv \frac{1}{e} \int_{-\infty}^{\infty} dt\, I_1(t).
\end{equation}
 
In the (adiabatic) limit of very slow pulses (large values of $\tau_P$ so that $\hbar/\tau_P$ is the smallest energy scale in the problem), the Landauer formula applies at all times,
\begin{equation}
    I_1(t) = \frac{e^2}{h} D_{\rm tot} V_0(t)
\end{equation}
and therefore,
\begin{equation}
    n_t =  D_{\rm tot} \int_{-\infty}^{\infty} dt \frac{eV_0(t)}{h}.
\end{equation}
The total number of electrons $\bar n$ sent by the pulse is ($D_{\rm tot}=1$),
\begin{equation}
    \bar n =  \int_{-\infty}^{\infty} dt \frac{eV_0(t)}{h}
\end{equation}
so that in the adiabatic limit we get the relatively boring result that
$n_t = D_{\rm tot} \bar n$. For our Gaussian pulse, $\bar n$ is simply given by,
\begin{equation}
    \bar n = \frac{1}{ 2\sqrt{\pi} } \frac{eV_P\tau_P}{\hbar}
\end{equation}
The situation becomes really interesting in the opposite limit of ultrafast pulses, the focus of this work, when $\tau_P$ is short with respect to the time of flight through the interferometer. However, we suppose that $\tau_P$ is still sufficiently long that the QPC can be properly described by their transmission at the Fermi energy i.e. that the energy dependence of the transmission can be neglected. In this limit, Ref.\cite{Gaury2014b} found that,
\begin{align}
\label{eq:shortpulses}
    n_{1/2} = \left[D_{\rm A }D_{\rm B } + (1-D_{\rm A })
 (1-D_{\rm B })\right] \bar n \pm \nonumber \\
 \sqrt{D_{\rm A }D_{\rm B } (1-D_{\rm A })
 (1-D_{\rm B })} \sin(\pi \bar n)\cos(\pi\bar n+\Phi) 
\end{align}
i.e. that in addition to the linear term, $n_1$ actually \emph{oscillates} with the amplitude $V_P$ of the pulse. Observing this prediction is the focus of some experimental activity as it would be a clear evidence of dynamical phase-coherence at the picosecond scale. If one interprets the two paths of the interferometer as a $|0\rangle$ and a $|1\rangle$, such experiments could become the basis of an electronic flying qubit where the QPCs play the role of a Hadamard gate and the phase $\Phi$ picked up in the loop is a rotation along the z axis \cite{Bauerle2018}. 

To understand the origin of Eq.\eqref{eq:shortpulses}, we need to describe what happens in the lead, at the place of the potential drop, before the interferometer. Before the pulse is applied, the electronic wavefunction of the edge state in the lead is a simple plane wave 
$\psi(y,t)= e^{iky-iEt}$ where $y$ is the one-dimensional coordinate along the edge state, $k$ the momentum and $E=E(k)$ the corresponding energy. Supposing that the voltage drop happens at $y=0$, during the duration of the pulse, the rear of the plane wave ($y<0$, inside the lead) acquires a phase
\begin{equation}
\label{eq:pulse_phase}
\phi(t) = \int_{-\infty}^t \frac{eV_0(u)}{\hbar}du
\end{equation}
with respect to the front ($y>0$ on the device side). The wave function thus becomes
\begin{equation}
\psi (y,t) = e^{iky -iEt -i\phi(t)\theta(-y)}
\end{equation}
where $\theta(y)$ is the Heaviside function. It follows that after the pulse is finished, there is a domain wall in the phase of the wavefunction between the front of the wave $y>0$ where $\psi(y,t)= e^{iky-iEt}$ and the rear of the wave $y<0$ where 
$\psi(y,t)= e^{iky-iEt -i\phi(+\infty)}$. The phase $\phi(+\infty)$ is
proportional to the number of electrons sent in the pulse 
$\phi(+\infty) = 2\pi\bar n$. The oscillations of Eq.\eqref{eq:shortpulses} should be understood as interferences between the front and the rear of the pulse as it propagates through the system.

\begin{figure}[H]
\centering{\includegraphics[width=\linewidth]{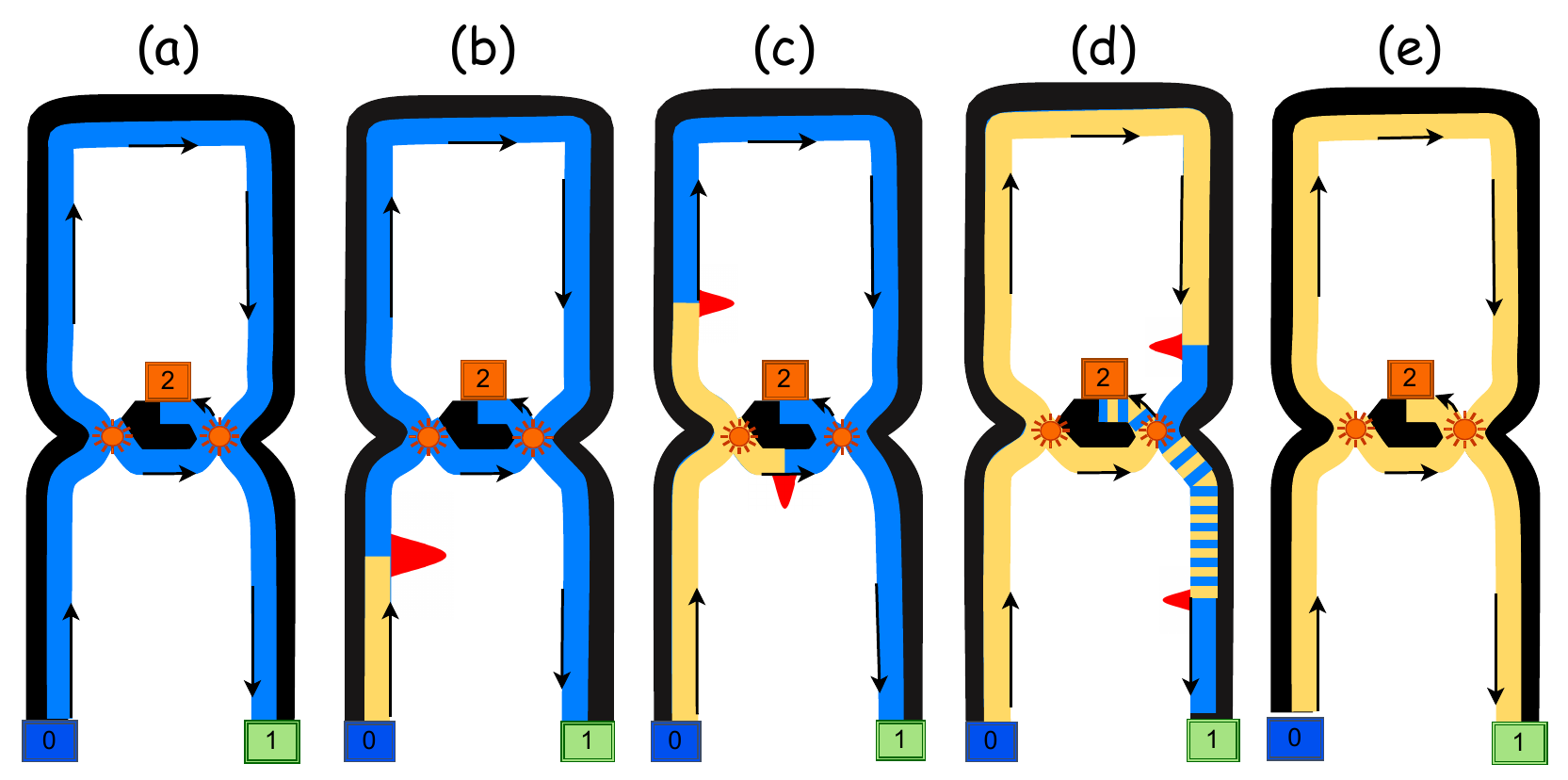}}
\caption{\label{fig:domainwall}Schematic representation of the dynamical 
interference effect. The color blue corresponds to the phase $e^{iky-iEt}$ where $y$ is the distance from contact $0$ following the edge. The color yellow corresponds to the phase
$e^{iky-iEt -i2\pi\bar n}$. (a) Initially, the edge state coming from contact 0 is at equilibrium, hence in blue. (b) The pulse arrives followed by its yellow tail.
(c) The pulse has been splitted by QPC A. (d) After partial recombination at QPC B of the lower part of the pulse, the system enters the transient period where the dynamical interference effect happens. The yellow-blue stripe corresponds to a region where the rear of the lower part of the pulse interferes with the front of the upper part. (e) After both parts of the pulse have reached contact $1$ and $2$, the system is back to equilibrium, albeit with a different global phase.
}
\end{figure}

More precisely, Fig.~\ref{fig:domainwall} shows a schematic of different snapshots of the phase during the propagation of the excitation generated by the pulse. Initially the phase is $e^{iky-iEt}$ everywhere
which is represented as blue in panel (a). In panel (b), the pulse has been generated and its rear carries an extra phase $2\pi\bar n$ shown in yellow. In panel (c) the pulse has been split by QPC A. The interesting situation occurs in panel (d) where the lower part of the pulse has crossed QPC B but the upper part has not reached it yet. At this moment, the rear of the pulse interferes with its front (yellow-blue stripe) giving rise to the above mentioned effect. Eventually, in panel (e), the pulse reaches all the electrodes and the phase becomes yellow everywhere. In summary, there is a transient range of time for which the lower pulse has reached the electrode $1$ but the upper part of the pulse has not done so. During this transient time, $I_1(t)$ contains an oscillating term $\cos(2\pi \bar n)$
that is the manifestation of the interference between the two phases in the front and rear of the pulse. The rest of this article is devoted to studying such an effect and MZI numerically and looking at the robustness of the above picture in presence of Coulomb repulsion.

%%%%%%%%%%%%%%%%%%%%%%%%%%%%%%%%%%%%%%%%%%%%%%%%%%%%%%%%%%%%%%%%%%%%%%%%%%%%%
\section{A time-dependent mean field treatment of interactions in a Mach-Zehnder Interferometer.}
\label{sec:model}
%%%%%%%%%%%%%%%%%%%%%%%%%%%%%%%%%%%%%%%%%%%%%%%%%%%%%%%%%%%%%%%%%%%%%%%%%%%%%

In this section, we describe our discrete model, our numerical technique and discuss the mean field approximation  used to treat electron-electron interactions.

\subsection{Model}
Let us start with the description of the discrete model used in the simulations.
Our starting point is a 2DEG (as typically found at an interface between GaAs and GaAlAs) described at the effective mass approximation level,
\begin{equation}
H_{\rm 2DEG} = \frac{1}{2m^*} \left( \mathbf{P} -e\mathbf{A}\right)^2 + V(\vec r).
\end{equation}
where, $m^*$ is the effective electron mass, $\mathbf{P}$ is the momentum vector, $e$ is the electronic charge, $\mathbf{A}$ is the vector potential and $V(\vec r)$ is the onsite potential on site $\vec r$. We discretize the problem on a rectangular grid with a lattice parameter $a$ using a five points stencil for the Laplacian (i.e. along one dimension, $\partial_{x}^2\psi \approx [\psi(x-a) -2\psi(x) +\psi(x+a)]/a^2$). The perpendicular magnetic field is described using Peierls 
substitution in the Landau gauge. The Coulomb interaction is treated through a contact Hubbard like interaction that we treat at the mean-field level. This is the self-consistent time dependent Hartree-Fock approximation which, we recall, is more accurate than its static counterpart. In particular, it captures the Random Phase Approximation (RPA) \cite{Negele2018}. We arrive at
\begin{equation}
\hat H =\sum_{ij,\sigma}
H_{ij}(t) c_{i\sigma}^{\dagger}c_{j\sigma}+
\sum_{i\sigma} V_i c_{i\sigma}^{\dagger}c_{i\sigma}+
\sum_{i\sigma} U_{i\sigma}(t) c_{i\sigma}^{\dagger}c_{i\sigma}
\end{equation}
where $i=(i_x,i_y)$ is a point in the grid and $c_{i\sigma}$ 
(resp. $c^\dagger_{i\sigma}$) is the usual fermionic operator that destroys a particle at site $i$ with spin $\sigma$ (resp. create). The external potential is $V_i \equiv V(\vec r=i_x a,i_y a)$. The mean-field interaction terms read,
\begin{equation}
U_{i\sigma} = U [\rho_{i\sigma}(t) - \rho_{i\sigma}(t=0)] 
\end{equation}
with the self-consistent condition,
\begin{equation}
\rho_{i{\bar{\sigma}}}(t) \equiv  \langle c_{i \bar \sigma}^{\dagger}(t)c_{i\bar \sigma}(t)\rangle.
\end{equation}
Here $\bar\sigma=\uparrow$ (resp. $\downarrow$) when $\sigma=\downarrow$ (resp. $\uparrow$).
In practice, we do not consider the role of the Zeeman field or exchange energy, so that the spin degree of freedom is purely spectator and from now on we drop the spin index. Note that
the system is infinite, i.e. the leads extend to infinity. The interaction and external potential terms are only present in the MZI region. We suppose that the external potential already accounts for possible self-consistent potential due to the static term, before one send the voltage pulse in the system.

The matrix elements $H_{ij}$ are zero except when $i$ and $j$ are nearest neighbours on the grid. In that case, according to the Landau gauge, we have
\begin{equation}
H_{ij} = \gamma e^{-i\phi_a (i_x + j_y)(i_y - j_y)/2}
\end{equation}
with $\gamma=\hbar^{2}/(2m^*a^2)$ and $\phi_a = Ba^2/(\hbar/e)$ the magnetic flux
through a $a \times a$ plaquette in units of the quantum of flux. At the interface between lead $0$ and the MZI, $H_{ij}$ acquires an extra phase $\phi(t)$ that corresponds to the voltage pulse,
\begin{equation}
\forall i\in \text{Lead 0}, 
\forall j\in \text{MZI}, 
H_{ij}\rightarrow H_{ij}e^{-i\phi(t)}.
\end{equation}

For our Gaussian voltage pulse, the acquired phase $\phi(t)$ is from Eq.(\ref{eq:pulse_phase}), where $V_{0}$ is defined using Eq.(\ref{eq:gauss_pulse}) gives,

\begin{equation}
\phi(t) = \frac{V_P\tau_P\sqrt{\pi}}{2}\left[1 + \text{erf}(t/\tau_P)\right]
\end{equation}

where, the voltage $V_P$ is measured in unit of $\gamma$ and the time $\tau_P$ in unit of $\hbar/\gamma$, setting $\hbar$ = 1. The external potential $V_i$ is zero except in the QPC regions where it smoothly interpolates between $V_i=V_A$ (resp. $V_B$) for $i$ at the center of QPC A (resp. QPC B)
to $V_i=0$ away from the QPC. The decay of the potential is chosen so that it is smooth with respect to the Fermi wave length $\lambda_F= 2\pi/k_F$. Our dispersion relation is
$E(\mathbf{k}) = 2\gamma \left[\cos (k_xa) +\cos (k_ya) \right]$.

Typical values used in the simulations are $\tau_P = 30/\gamma$ and $V_P=0.023\gamma$
which corresponds to $\bar n= 0.2$ electrons per pulse. The Fermi energy is set at
$E_F/\gamma= -3.2$ which corresponds to $\lambda_F/a=2\pi/\sqrt{E_F/\gamma + 4} \approx 7$
and a (spinless) electronic density per lattice site $n_s a^2 = (E_F/\gamma +4)/(4\pi) \approx 0.064$.
We use $\phi_a = 0.44$ which corresponds to a magnetic length 
$l_B/a= 1/\sqrt{\phi_a} \approx 1.51$ and a cyclotron energy
$\hbar\omega_c/\gamma = 2\phi_a \approx 0.88$.

For a discretization parameter $a=13$ nm, the simulation corresponds to a sample of
realistic size $2.6\mu$m $\times 1.3 \mu$m. The width is limited to $W= 130$ nm for efficiency purpose but this is sufficient for the two counter propagating edge states to propagate freely without unwanted  back scattering. We get $\gamma = 3.3 $ meV and the electronic density corresponds to $n_s = 0.75 \ 10^{15} m^{-2}$, a typical value for GaAs/GaAlAs 2DEGs (on the low side). It corresponds to $\lambda_F \sim 91\text{ nm}$. The externally applied magnetic field is $B=1.71$ T which corresponds to magnetic length
$l_{B}=19$ nm and a cyclotron frequency of $E_{c}=\hbar\omega_{c}=2.91$
meV.

To estimate the value of $U$ that would be relevant, we need to understand how the screening
takes place in the sample since we only use a local contact interaction while the actual Coulomb repulsion is long range. A particularly easy situation is when this screening is induced \emph{externally} by adding a second 2DEG at a distance $d$ below the one of interest in order to force this screening and reduce the role of the electron-electron interactions. Such a scheme has been implemented successfully in a set of recent experiments \cite{Nakamura2020,Nakamura2022} in order to suppress Coulomb blockade and observe the phase associated with abelian anyons. In such a system, one can model the interaction by a simple planar capacitance $\epsilon/d$ per unit surface, i.e. the addition of $\delta \rho$ electron on
a site gives rise to a raise of on site energy $\delta U =\epsilon a^2/(de^2)\delta \rho$. Making contact with our model, we arrive at,
\begin{equation}
U = \frac{2m^* e^2 d}{\epsilon\hbar^2}
\end{equation}   
for $d= 10$ nm, a typical experimental value, we obtain $U\approx 25$ which gives us the order of magnitude of $U$ we should have in mind.

\begin{figure}[H]
\centering\includegraphics[width=\linewidth]{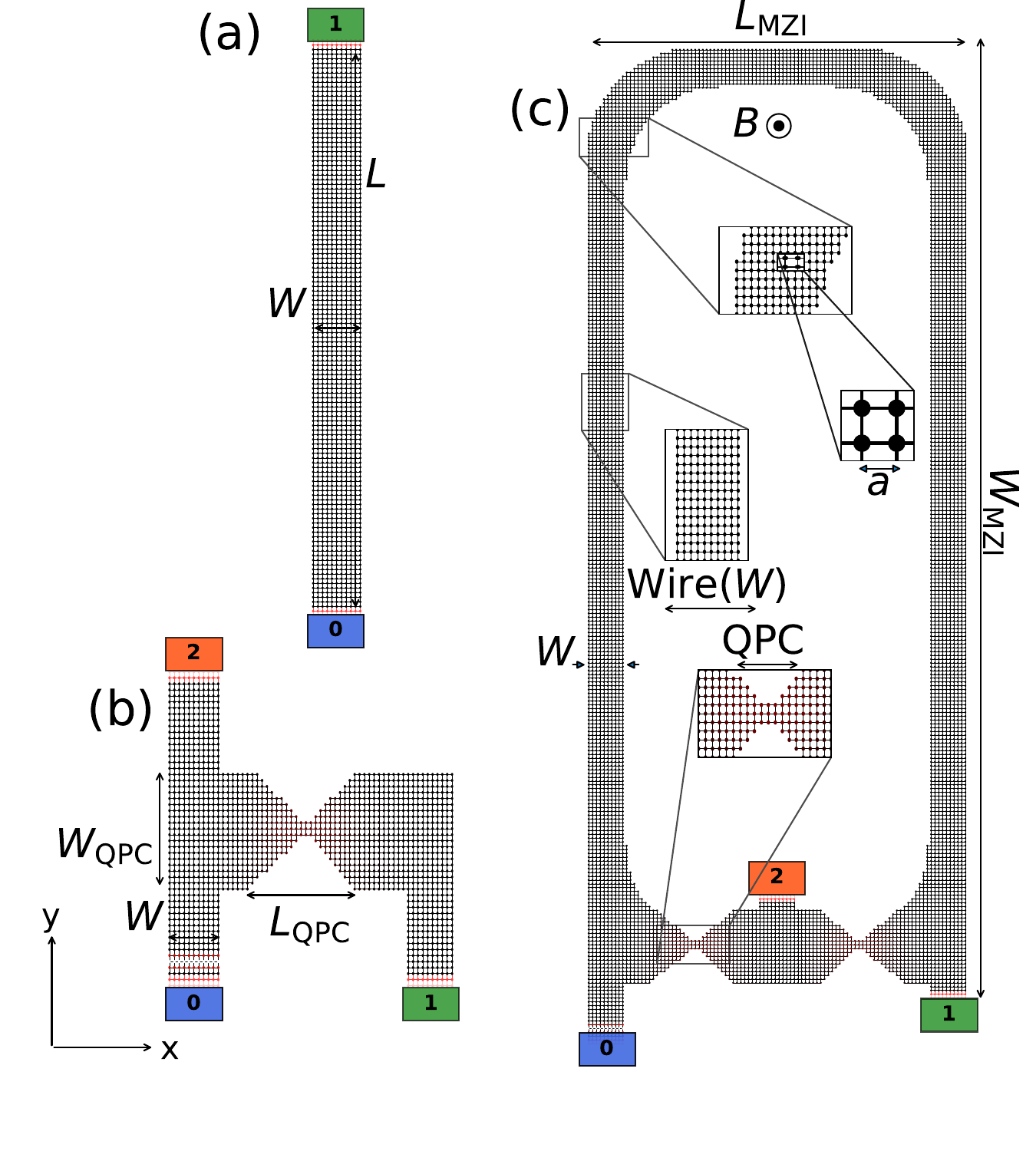}
\caption{\label{fig:model} Tight-binding models considered in this work.
(a) A quasi-one dimensional wire connected to two semi-infinite electrodes 0 and 1.
Width: $W=10$ sites, length: $L=200$.
(b) a Quantum Point Contact (QPC) connected to three electrodes 0, 1 and 2.
$L_{\text{QPC}}=W_{\text{QPC}}=22$.
(c) the Mach-Zehnder interferometer (MZI) that contains the QPC and the wire region as subparts. $L_{\text{MZI}}=100$ and $W_{\text{MZI}}=250$ for a total of around 
$17 000$ sites. The different insets are zoom into different regions. 
}
\end{figure}
 
%%%%%%%%%%%%%%%%%%%%%%%%%%%%%%%%%%%%%%%%%%%%%%%%%%%%%%%%%%%%%%%%%%%%%%%%%%%%%
\subsection{Numerical Technique}
%%%%%%%%%%%%%%%%%%%%%%%%%%%%%%%%%%%%%%%%%%%%%%%%%%%%%%%%%%%%%%%%%%%%%%%%%%%%%

To solve the model, we use the self-consistent module of the Tkwant package \cite{Kloss2021,Tkwant2021} that has been developed by some of us. We refer to the Tkwant documentation for details of how the system of equations  is solved. First, the scattering states $\Psi_{\alpha E}$ of the problems are calculated (before the pulse is applied) using the Kwant package \cite{Groth2014} which itself uses 
the technique described in \cite{Xavier2024}.
\begin{equation}
\sum_j [H_{ij} \Psi_{\alpha E}(j)] +V_i \Psi_{\alpha E}(i)= E \Psi_{\alpha E}(i)
\end{equation}
for all open modes $\alpha$ and $N_E$ different values of energy that span all the  
occupied part of the lead bands $E\in [-4\gamma,E_F/\gamma]$ (here a single band is occupied as we work within the lowest Landau level, see the section on the wire).
Second, we solve $N_E$ Schrödinger equations self-consistently in parallel,
\begin{multline}
i\partial_t \Psi_{\alpha E}(i,t) =
\sum_j \bigl[ H_{ij} \Psi_{\alpha E}(j,t) \bigr]
+ V_i \Psi_{\alpha E}(i,t) \\
+ U_i(t)\, \Psi_{\alpha E}(i,t)
\end{multline}
where $\Psi_{\alpha E}(i,t=0)=\Psi_{\alpha E}(i)$ and $U_i(t) = U [\rho_i(t) - \rho_i(0)]$
with
\begin{equation}
\rho_i(t) =  \sum_\alpha \int \frac{dE}{2\pi} 
|\Psi_{\alpha E}(i,t)|^2  f_\alpha(E)
\end{equation}
where $f_\alpha(E)$ is the Fermi function in lead $\alpha$ (here we work at zero temperature and all leads share the same Fermi energy). Tkwant solves this problem using a prediction-correction scheme \cite{Kloss2021} that takes advantage of the fact that $U_i(t)$ has
a much slower evolution than $\Psi_{\alpha E}(i,t)$.

A typical simulation shown in this manuscript corresponds to the self-consistent solution of
$N_E \approx 500$ Schrödinger equations in parallel, each of them containing a total of
$17050$ lattice sites for the full Mach-Zehnder interferometer.
 
%%%%%%%%%%%%%%%%%%%%%%%%%%%%%%%%%%%%%%%%%%%%%%%%%%%%%%%%%%%%%%%%%%%%%%%%%%%%%
\subsection{A comment on current conservation and gauge invariance}
%%%%%%%%%%%%%%%%%%%%%%%%%%%%%%%%%%%%%%%%%%%%%%%%%%%%%%%%%%%%%%%%%%%%%%%%%%%%%

In this section, we comment on the importance of treating electron-electron interaction
at \emph{some} level when dealing with time-dependent phenomena, even when the system under consideration, here a relatively high density 2DEG, is weakly correlated. In a 2DEG, the strength of electron-electron interactions is  measured by the dimensionless parameter $r_s=m^*e^2/(4\pi\epsilon\hbar^2\sqrt{\pi n_s})$ which is the ratio of the typical electrostatic energy over the kinetic energy. The system starts to be correlated for $r_s>10$ with the transition to the Wigner crystal around $r_s=30$. With our parameters $r_s\approx 2$ which is well within the Fermi liquid regime and typical of a GaAs/GaAlAs heterostructure.

Yet, one should remember that at equilibrium the system is essentially neutral. As soon as one starts to, say, send a voltage pulse inside the system some extra charge will propagate within the 2DEG carrying an important amount of electrostatic energy. This charge will be screened in one way or another either by the 2DEG itself or by surrounding metallic elements in e.g. the electrostatic gates. Let us consider the charge conservation equation,
\begin{equation}{
\frac{\partial \rho_i}{\partial t} = \sum_{j} I_{ij}}
\end{equation}
where $\rho_i = \langle c^\dagger_{i}c_{i}\rangle$ is the charge on 
site $i$ and $I_{ij} = -2\text{Im} [ H_{ij} \langle c^\dagger_{i}c_{j}\rangle]$ is the current going from site $j$ to site $i$. This current term is the discrete version of the divergence of a current density 
$\nabla \cdot \textbf{I}$.
Remembering Gauss law, the appearance of the term $\partial \rho_i/\partial t$ must correspond to the appearance of an electric field $\textbf{E}$ with 
\begin{equation}
\nabla \cdot \frac{\partial \textbf{E}}{\partial t} = \frac{e}{\epsilon}
\frac{\partial \rho}{\partial t}
\end{equation}
i.e. a displacement current $\partial \textbf{E}/\partial t$ adds on to the particle current $\textbf{I}\rightarrow \textbf{I} + \partial \textbf{E}/\partial t$ in order to screen the charge. This displacement current is the key ingredient that is missing in the non-interacting theory.

A dual way to look at the same problem is through gauge invariance. Suppose that one raises the electric potential of the system \emph{everywhere} in the system (i.e. in the lead 
\emph{and} in the MZI) by a pulse $V(t)$. Then it is trivial to observe that this pulse can be ``gauged out" by redefining the wave function as 
\begin{equation}
\psi_{\alpha E}(i,t) = e^{-i\int_0^t du V(u)} \bar \psi_{\alpha E}(i,t)
\end{equation}
where $\bar \psi_{\alpha E}(i,t)$ obeys the Schrödinger equation without the time dependent term. In other word if one send a voltage pulse everywhere, it does nothing. In the non-interacting version of our model however, we have supposed implicitly that the electric potential of the MZI is fixed as if the potential difference (in an actual experiment there are only potential \emph{differences}) could be applied between the electrode and the MZI itself. In reality, however, this potential can only be applied between different electrodes or between electrodes and electrostatic gates. It follows that the electric potential inside the MZI must be allowed to vary in time to restore the correct physical behavior.

In short, in order to get a physically coherent picture, the electrostatic of the problem must be respected. In a discrete model this amounts to introduce a capacitance matrix 
$C_{ij}$ with $\sum_j C_{ij} U_j(t) = \rho_i(t)$. The self-consistent time dependent Hartree-Fock approximation used in this work corresponds to $C_{ij}=U\delta_{ij}$.
The locality of the matrix emerges spontaneously from the response of the 2DEG but can also be helped by the presence of nearby metallic gates (see Fig.4 in \cite{Armagnat2019}).
Previous work have studied the effect of the electron-electron interactions, at this level of approximation, inside a wire geometry and in the absence of magnetic field. It was found in particular that, as expected by bosonisation and Luttinger theory, the Fermi velocity gets renormalized into a (larger) Plasmon velocity. The calculated plasmon velocity agreed
quantitatively with what is obtained from (multichannel) Luttinger theory and a quantitative match, in the absence of fitting parameter, was found with experiments measuring this velocity. The present work extend these works to the quantum Hall regime and to an actual full fledged MZI interferometer.

%%%%%%%%%%%%%%%%%%%%%%%%%%%%%%%%%%%%%%%%%%%%%%%%%%%%%%%%%%%%%%%%%%%%%%%%%%%%%
\section{Propagation of a pulse in a quasi-one dimensional geometry in the lowest Landau level}\label{sec:1D-Wire}
%%%%%%%%%%%%%%%%%%%%%%%%%%%%%%%%%%%%%%%%%%%%%%%%%%%%%%%%%%%%%%%%%%%%%%%%%%%%%
We start presenting our numerical results with a simple ``wire" geometry corresponding
to Fig. \ref{fig:system}(a). The goal is to characterize the regime in which the simulations are performed and isolate the main effect of the interactions that will take place in the Mach-Zehnder interferometer, the renormalization of the velocity.

\subsection{Characterization of the DC physics}
\begin{figure}[H]
\includegraphics[width=0.9\linewidth, height=1.1\linewidth]{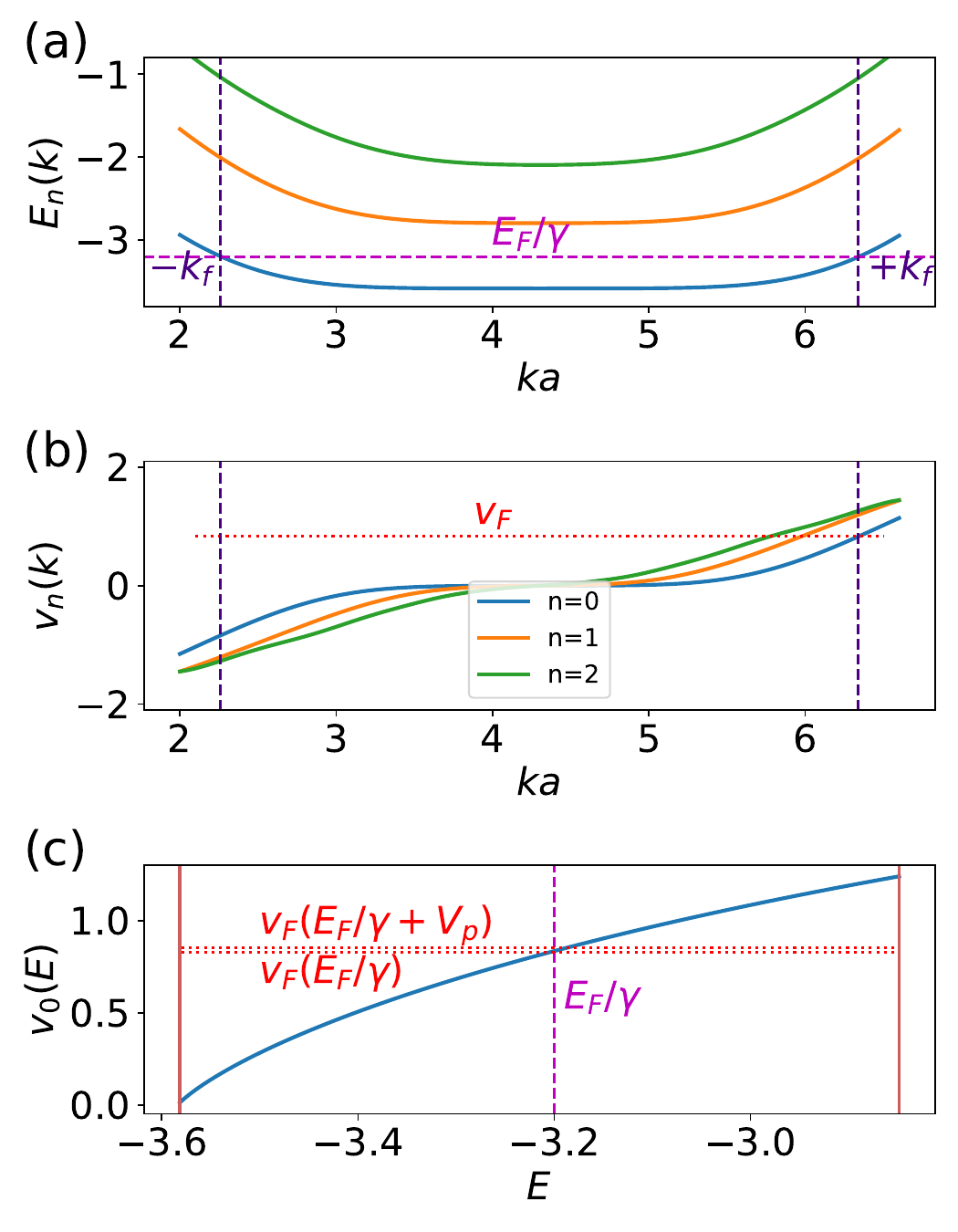}
\caption{\label{fig:wire_bands}
DC characterization of the wire model at $\phi_a=0.44$: dispersion relation.
(a) Band dispersion $E_n(k)$ of the infinite electrodes for the three lowest Landau levels. The horizontal dashed line marks the Fermi level $E_F/\gamma$.
(b) The corresponding band dispersion velocity $v_n(k)$.
(c) Same as (b) but versus energy for the lowest Landau level $v_0(E)$. The left and the right red bars are respectively the bottom of the lowest and first Landau levels.}
\end{figure}

We start with characterizing our wire before we send the pulse. Fig. \ref{fig:wire_bands}(a) shows the
dispersion relation $E_n(k)$ of the wire. We observes the usual Landau levels. We place the Fermi level so that a single chiral edge state is occupied. Fig. \ref{fig:wire_bands}(b) show the velocity
$v_n(k)$ for the same bands. Formally, $v_n(k) = (1/\hbar) dE_n/dk$ but here the velocity is calculated directly using the eigenstate of the current operator \cite{Xavier2024}. A plot of the velocity $v_0(E)$ from the bottom of the lowest Landau level to the next one is shown in Fig. \ref{fig:wire_bands}(c) highlighting the velocity at the Fermi level $E_F/\gamma$ and the velocity at $E_F/\gamma+V_P$. We see that $V_P$ is sufficiently small in these numerics that the velocity can be effectively considered as constant (even though in the numerics its full energy dependence is fully taken into account).

\begin{figure}[H]
\centering\includegraphics[width=\linewidth]{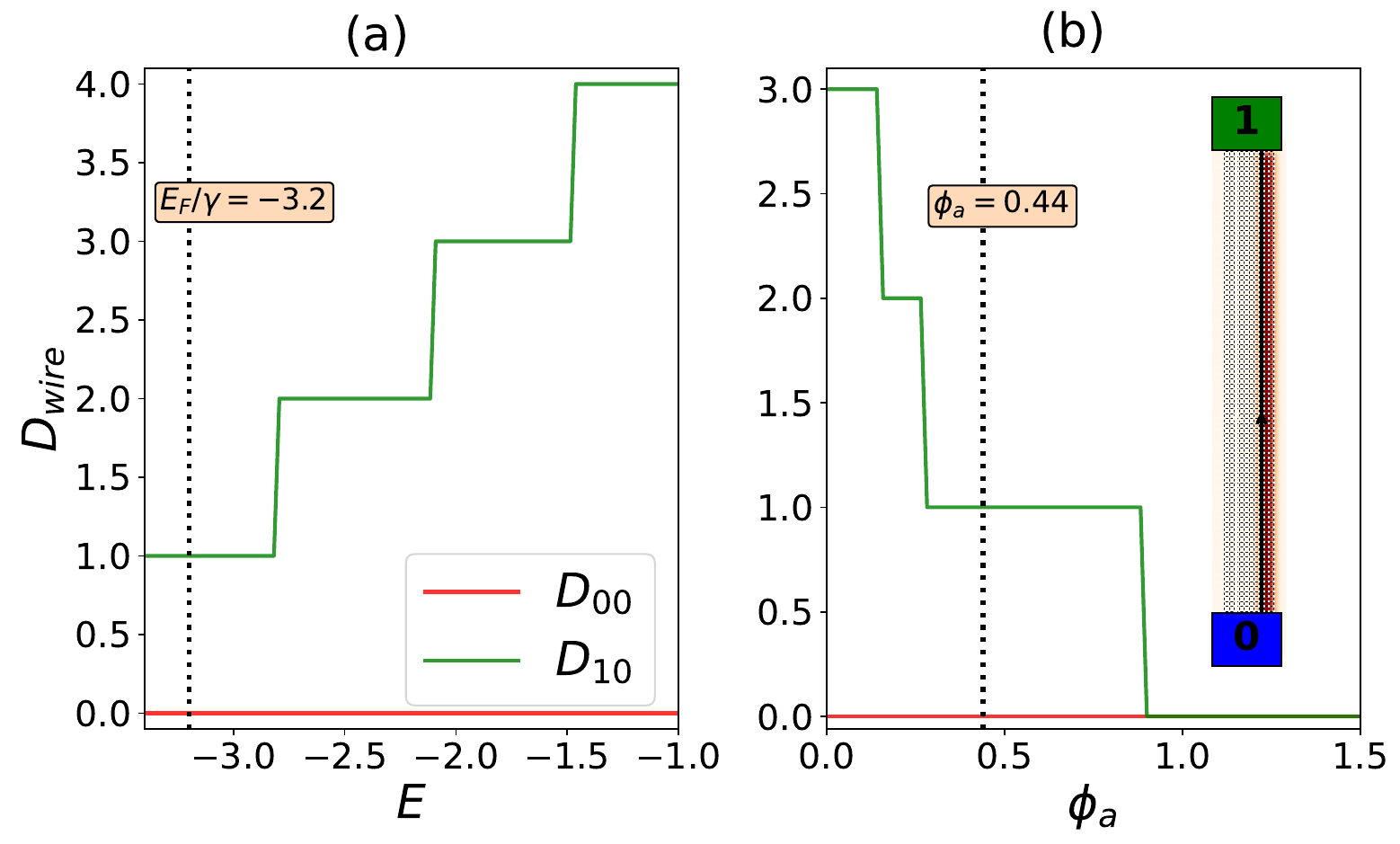}
\caption{\label{fig:wire_DC_characteristics} DC characteristics of the wire model: transmission. 
(a) Transmission (green) and reflection (red, vanishing) probability of the wire versus energy at $\phi_a=0.44$. 
(b) Transmission and reflection probability versus magnetic field $\phi_a$ at $E = E_F/\gamma=-3.2$. Inset: Colormap of the local current density injected from contact $0$ at $E = E_F/\gamma=-3.2$ and $\phi_a=0.44$. One clearly observes that the propagation takes place on the right edge.}
\end{figure}

The transmission $D_{10}$ and reflection $D_{00}$ probabilities versus energy (panel a)
and magnetic field (panel b) are shown respectively in Fig. \ref{fig:wire_DC_characteristics}. The sample is invariant under translation which imposes $D_{00}=0$. The Fermi energy is placed such that $E_F/\gamma$ is away from both the opening of the second channel and the bottom of the band, even in presence of the voltage pulse. Panel b shows the transmission versus magnetic field; the length of the last plateau is an indication of the system entering the quantum Hall regime. The position of the $n^\text{th}$ Landau level is expected theoretically to be at $E_F/\gamma =-4 + 2n\phi_a$ which indeed matches our observation ($\phi_a\approx 0.4$). A snapshot of the corresponding current density is shown as an inset of panel b, confirming that it is indeed an edge state.

\subsection{Pulse propagation}
\label{sec:pulse_vel}

\begin{figure}[H]
\includegraphics[height=\linewidth, angle=90]{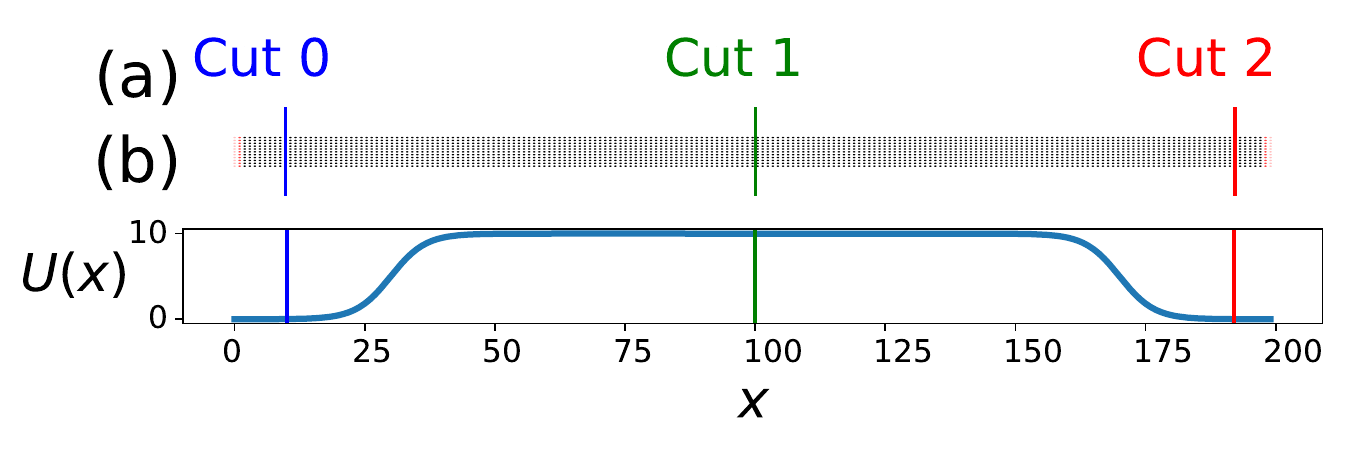}
\includegraphics[width=\linewidth]{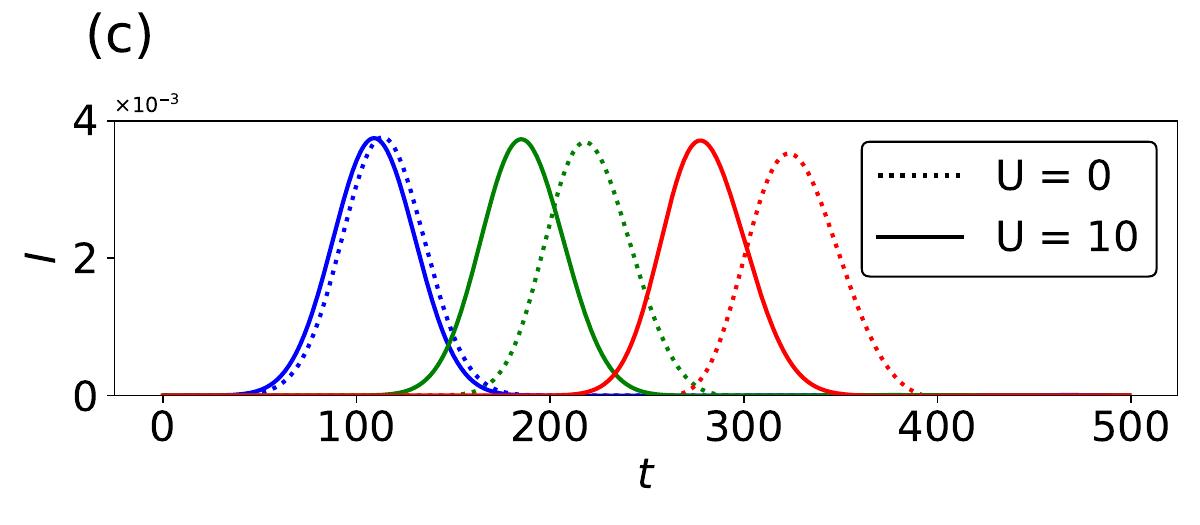}
\caption{\label{fig:wire_AC_characteristics}
Propagation of a voltage pulse inside the wire model \ref{sec:1D-Wire}. 
(a) Schematic of the wire with the three positions where the current is recorded.
(b) Profile of the interaction $U(x)$: in order to avoid backscattering due to an abrupt switch on of the interaction, the interaction is switched on adiabatically over a length of  $200$ sites.
(c) Current versus time $t$ at the three positions shown in (a) (respectively blue, green, red) with (full lines, $U=10$) and without (dotted lines, $U=0$) interaction.}
\end{figure}

Having characterized the wire, we proceed with sending a voltage pulse through it 
and studying the propagation of the associated excitation. To avoid spurious reflections created by an abrupt switch on of the interaction, the interaction is adiabatically switched on in the wire over a length of $\sim 20 a$, as shown in Fig. \ref{fig:wire_AC_characteristics}(b). More precisely, we replace the constant $U$ with a spatially varying $U(x)$ that vanishes in the lead and saturates at the value $U$ in the wire, following \cite{Kloss2018}. The current versus time is measured at three cross sections of the wire (labeled cut 0, 1 and 2 in Fig. \ref{fig:wire_AC_characteristics}(a) and the results are shown in Fig.  \ref{fig:wire_AC_characteristics}(c). The peaks at the different cuts appear at different times allowing one to follow precisely the propagation of the pulse in the wire,
hence to extract a velocity. One observes that the propagation is faster in presence of interaction ($U=10$, full lines) compared to the non-interacting case ($U=0$, dotted lines)
which corresponds to the renormalized velocity characteristic of Luttinger liquids.

\begin{figure}[H]
\centering\includegraphics[height=\linewidth,angle=-90]{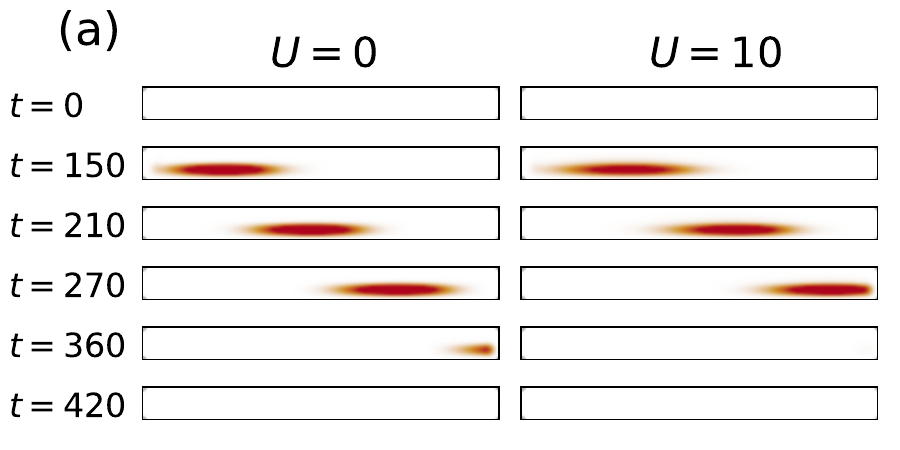}
\centering\includegraphics[width=\linewidth]{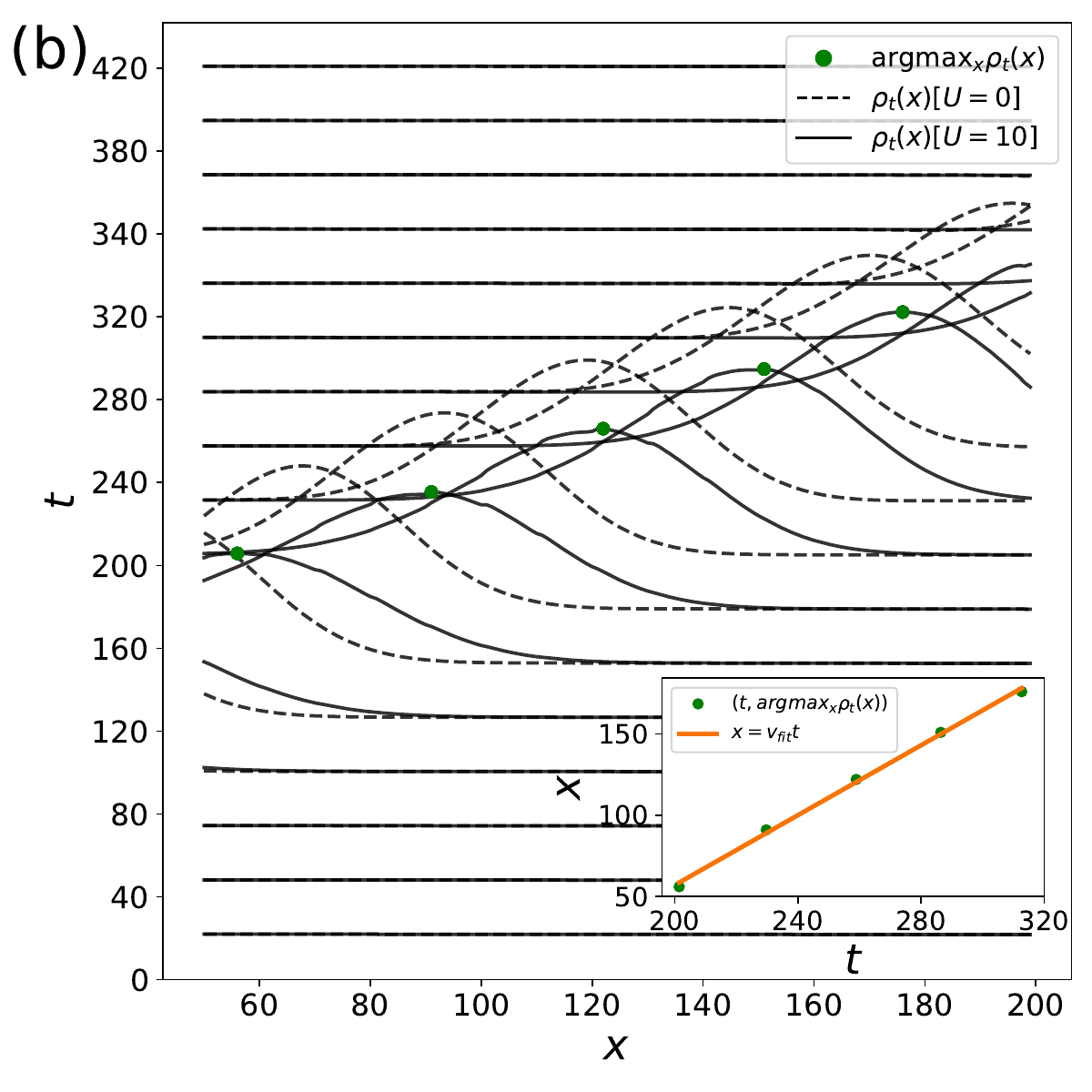}
\caption{\label{fig:wire_pulse}
Propagation of a voltage pulse through the wire: electronic density.
(a) Colormap of the density $\rho(x,y)$ at different times with (right) and without (left)  interaction. 
(b) Same data but the density is integrated across the transverse direction $y$ without 
(dashed) and with (full line, $U=10$) interaction.
The different curves $\rho_t(x)=\sum_y \rho(x,y,t)$ are translated vertically by a shift proportional to the time $t$. 
The inset shows the position of the maximum $x = \text{argmax}_x \rho_t(x)$ versus $t$ for the different curves shown in the main plot and $U=10$. The full line is a linear fit $x=v_\text{fit} t$ from which the velocity of the pulse is extracted.}
\end{figure}

\begin{figure}[H]
\centering\includegraphics[width=\linewidth]{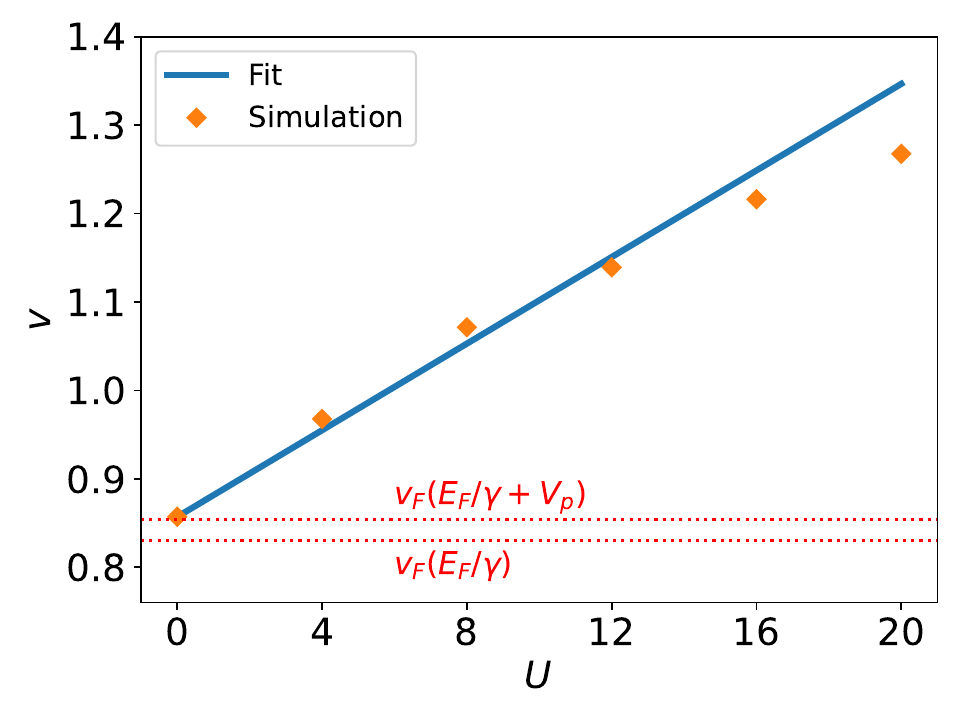}
\caption{\label{fig:pulse_vel_in_interaction} Pulse velocity $v$ extracted from the simulation (see Fig. \ref{fig:wire_pulse}(b) ) versus interaction strength $U$. 
The horizontal red dashed lines mark the window of non-interacting velocity
Fermi velocity $v(E_F/\gamma)\le v \le v(E_F/\gamma+V_P)$.
The blue full line corresponds to Eq. \eqref{eq:pulse_vel} with $l=6.5$.}
\end{figure}

The propagation can also be monitored at the level of the electronic density,
as shown in Fig. \ref{fig:wire_pulse}.  Fig. \ref{fig:wire_pulse}(a) shows a few snapshots of the excess electronic density
without (left) and with (right) interaction. We observe again an increase of the velocity
in presence of interactions. To measure this effective velocity quantitatively, we record the position of the maximum $x_i$ in space of the electronic density at different times $t_i$ (see Fig. \ref{fig:wire_pulse}(b)); the slope $x_i = v t_i$ is the renormalized velocity. The resulting
curve $v(U)$ is shown in Fig. \ref{fig:pulse_vel_in_interaction}. As expected, $v(U=0)$ is given by the non-interacting
velocity provided by the band structure.

To understand the effect of interaction on the velocity, we resort to the semi-classical theory developed in \cite{Kloss2018}. Qualitatively, the excess density due to the pulse gives rise to an associated gradient of potential, hence a force, which self-consistently accelerates the pulse. This effect can be accounted for quantitatively at the Boltzmann equation level which has been found to (i) recover exactly the Luttinger liquid expression derived from bosonization theory \cite{Kloss2018} and (ii)
accounts quantitatively for the experimental data of \cite{Roussely2018} (using a proper electrostatic solver instead of the simple contact interaction used in the present work).
With respect to \cite{Kloss2018}, we have two modifications to operate: first, 
the edge state is chiral so we have only forward propagation to account for instead of the
forward and backward densities. Second, the edge state has a finite width $l$ instead of the purely one-dimensional case of \cite{Kloss2018}. One arrives at a Liouville equation of the form
\begin{equation}
\partial_t \rho(x,t) + \left(v_F + \frac{U}{2\pi l} \right) \partial_x \rho(x,t) = 0
\end{equation}
so that the velocity is given by a simple linear form
\begin{equation}
\label{eq:pulse_vel}
v = v_F + \frac{U}{2\pi l}.
\end{equation}
Here $l \propto l_B$ is the width of the edge state which we measure to be of the order of
$l \approx 5$ on the density profile of Fig. \ref{fig:wire_pulse}. The best fit to the numerical data is given 
by $l\approx 6.5$ and we also note that the numerical data is not perfectly linear. 

%%%%%%%%%%%%%%%%%%%%%%%%%%%%%%%%%%%%%%%%%%%%%%%%%%%%%%%%%%%%%%%%%%%%%%%%%%%%%
\section{Characterization of the Quantum Point Contact (QPC).}
\begin{figure}[H]
\includegraphics[width=0.48\linewidth]{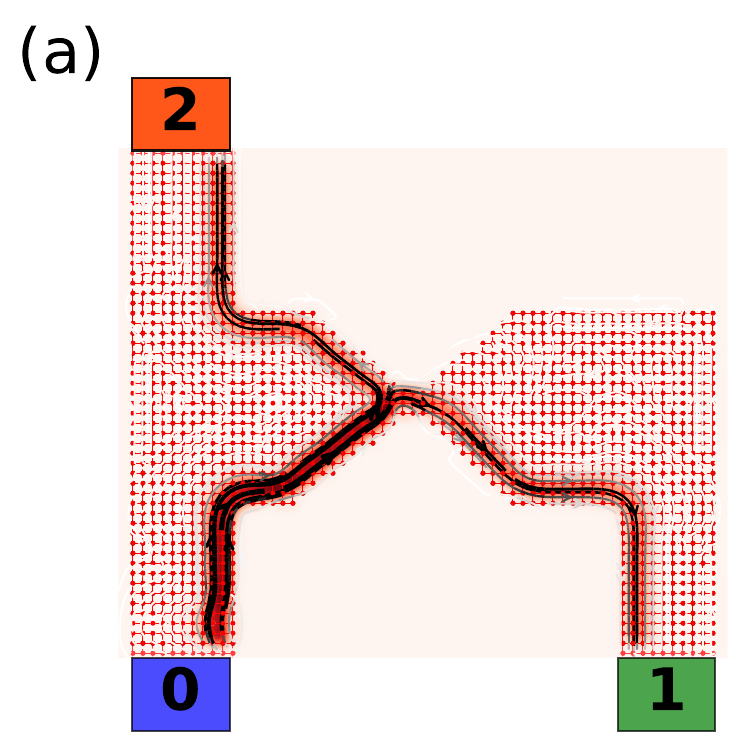}\includegraphics[width=0.48\linewidth]{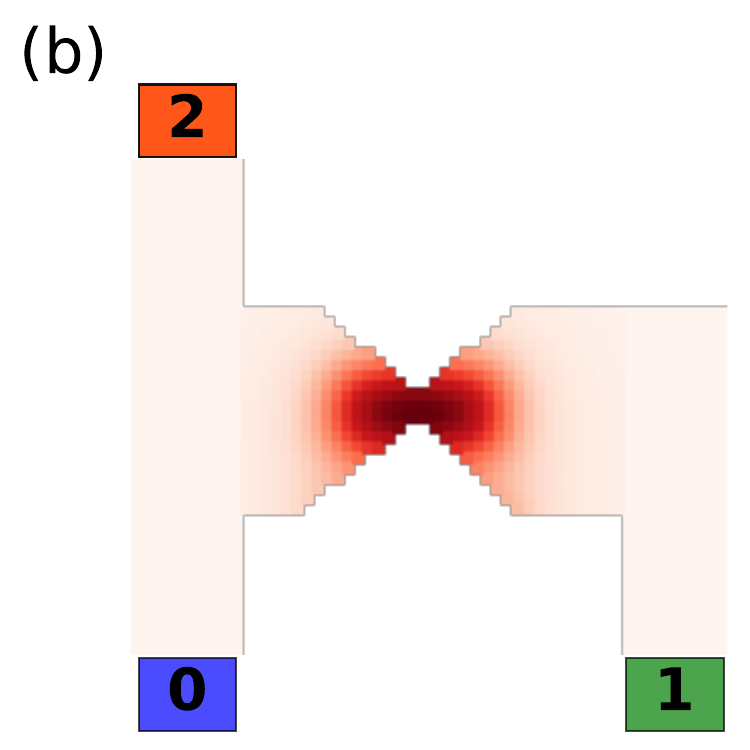}\\ \includegraphics[width=\linewidth]{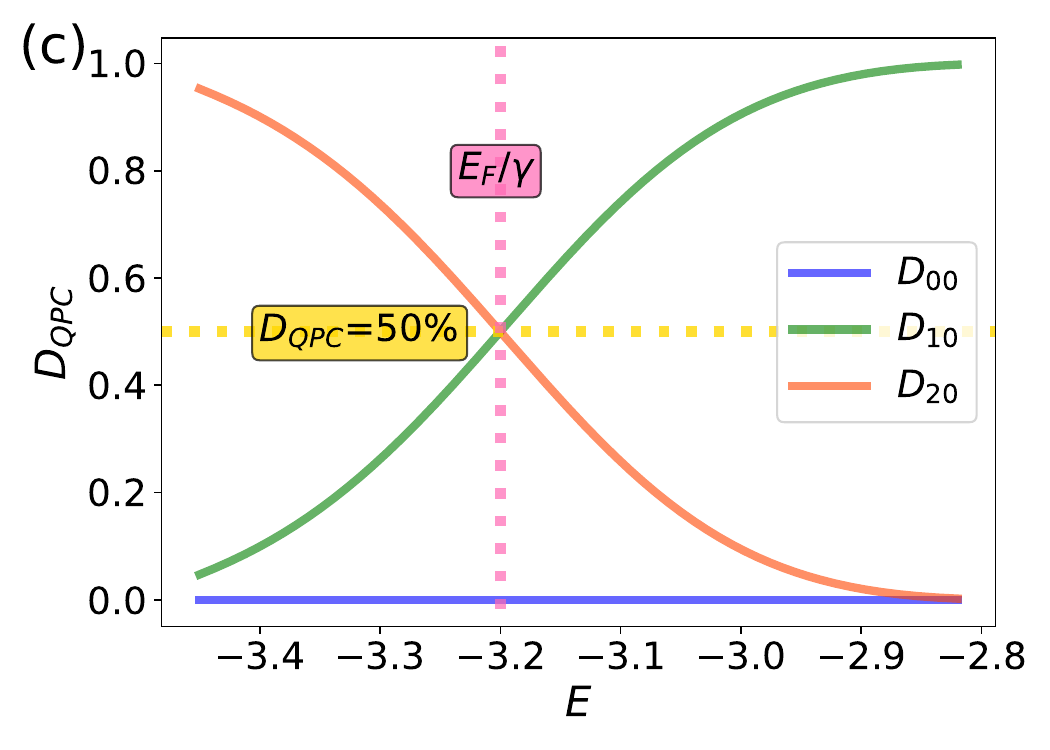}
\caption{\label{fig:QPC_DC_characteristics}
DC characteristics of the QPC model at $\phi_a=0.44$. 
(a) Colormap of the Current density in the QPC injected from contact $0$. 
(b) Colormap of the potential in the QPC. The potential is maximal with value $V_g$ at the center of the QPC and decays smoothly away from it; 
(c) Transmission from contact $0$ to the three contacts $0$ (blue), $1$ (green) and $2$ (orange) versus energy $E$. The gate potential $V_g$ has been tuned so that the QPC works as a beam splitter $D_{10}=D_{20}=1/2$ at the Fermi energy $E_F/\gamma=-3.2$.}
\end{figure}

%%%%%%%%%%%%%%%%%%%%%%%%%%%%%%%%%%%%%%%%%%%%%%%%%%%%%%%%%%%%%%%%%%%%%%%%%%%%%
\label{sec:QPC}
We now turn to the characterization of the QPC. A smooth (with respect to $\lambda_F$) potential $V_i$ is added in the QPC region so that $V_i=0$ away from the center of the QPC 
and reaches a maximum $V_i=V_g$ in the center of the constriction, see Fig. \ref{fig:QPC_DC_characteristics}(b) for a typical potential used in the numerics. The gate potential $V_g$ is used to tune the QPC to be a perfect beam splitter of transmission $D_\text{QPC}=1/2$ at the Fermi level. Fig. \ref{fig:QPC_DC_characteristics}(c) shows the resulting dispersion curve for the QPC transmission versus energy and Fig. \ref{fig:QPC_DC_characteristics}(a) shows a snapshot of the edge state partition at the Fermi level.

\begin{figure}[H]
\includegraphics[width=\linewidth]{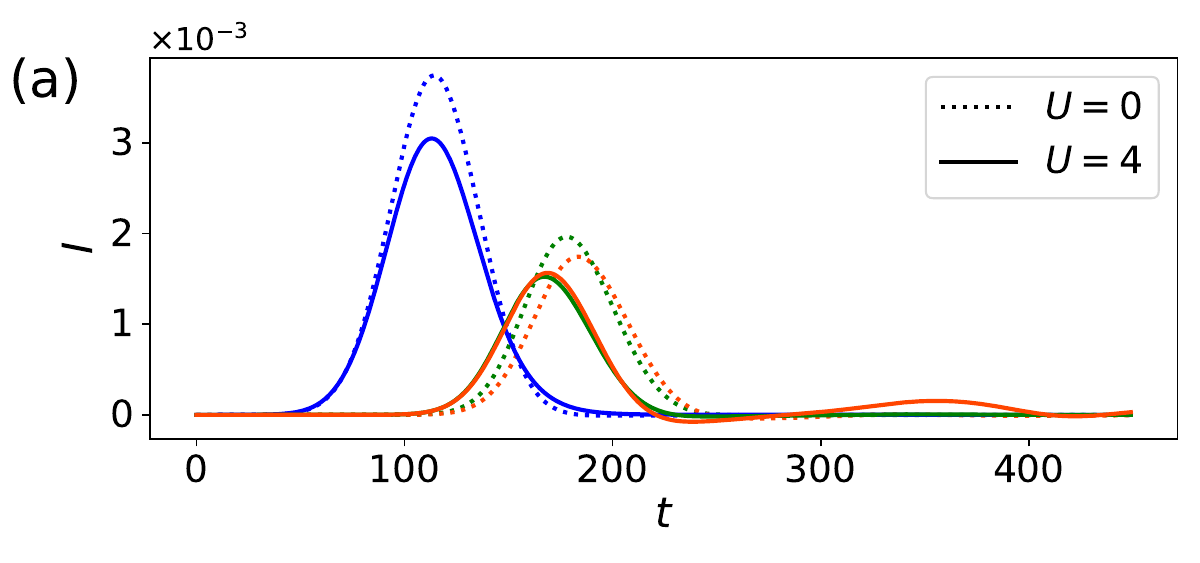}
\includegraphics[width=\linewidth]{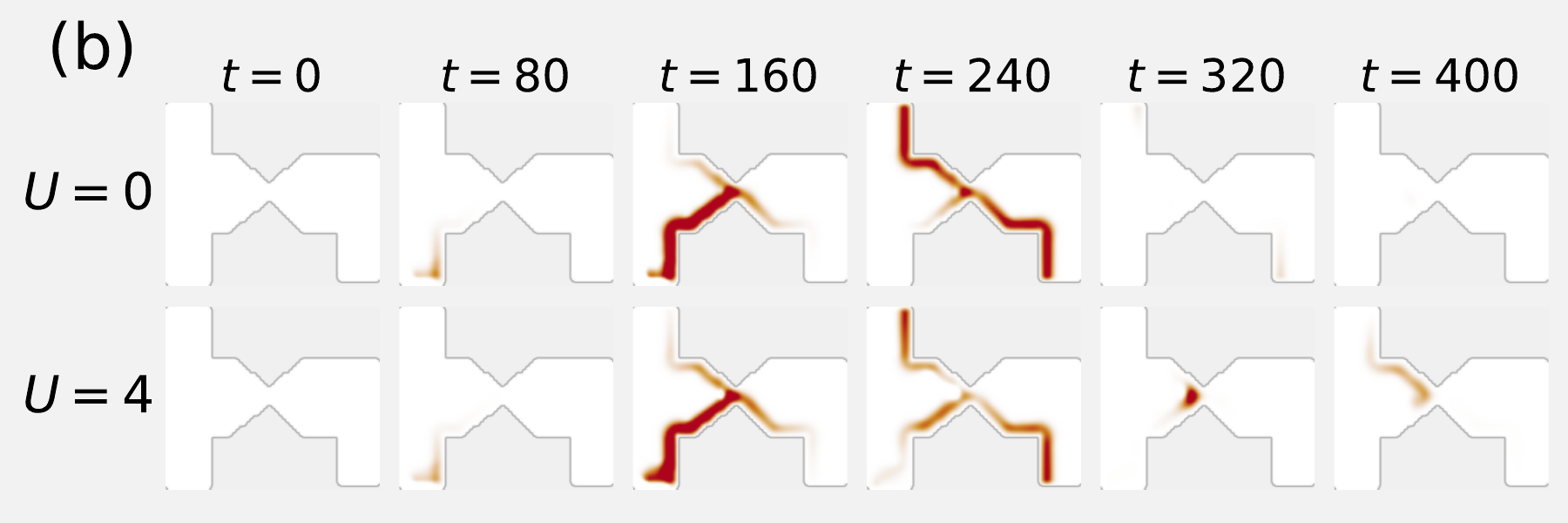}
\caption{\label{fig:QPC_AC_characteristics}
Pulse injection inside the QPC model.
(a) Current $I(t)$ measured at contact $0$ (blue), $1$ (green) and $2$ (red)
 without ($U=0$, dotted line) and with interaction ($U=4$, full line) after injection of a pulse at contact $0$.
(b) Colormap of the electronic density at various times without (upper panels) and with (lower panels) interaction.}
\end{figure}

The current versus time at the three contacts is shown in Fig. \ref{fig:QPC_AC_characteristics}(a) while Fig. \ref{fig:QPC_AC_characteristics}(b) shows the corresponding snapshots of the excess density at various times. Without interaction, the pulse is cleanly split between the transmitted part to contact 1 and 2 indicating that the pulse amplitude is small enough for the non-linearities of the QPC to play a minor role.
However, when one switches on the interaction, we observe the appearance of a small new maxima in the reflected current around $t=350$ (Fig. \ref{fig:QPC_AC_characteristics}(a), thin red line) which can also
be seen in the lingering density in Fig. \ref{fig:QPC_AC_characteristics}(b) panel $t=320$. The fact that interaction effects are enhanced in the QPC region is to be expected: close to the constriction, the local Fermi energy $E_F-V_g$ is much smaller (in fact essentially vanishes since we work in the $D_\text{QPC}=1/2$ regime) while the density of excess charge is essentially unaffected (due to charge conservation). Hence, the electrons are very slow at the constriction which makes the effective strength of interaction (the "local" $r_s$ parameter, so to speak) with respect to kinetic energy much larger; non-linear effects are expected to appear. This increase of the effect of electron-electron interactions is believed to be the origin of the 0.7 anomaly observed in many different QPC devices \cite{Micolich2011}, although this phenomena is probably not captured at the level of approximation we work with. In the present numerics, this results in the mild alteration of the shape of the pulse visible in 
Fig. \ref{fig:QPC_AC_characteristics}. Besides the small secondary peak, we observe a slight decrease of the width of the pulse as interaction is switched on.

\section{Simulations of the full Mach-Zehnder interferometer}
\label{sec:MZI}
\begin{figure}[H]
\includegraphics[width=\linewidth]{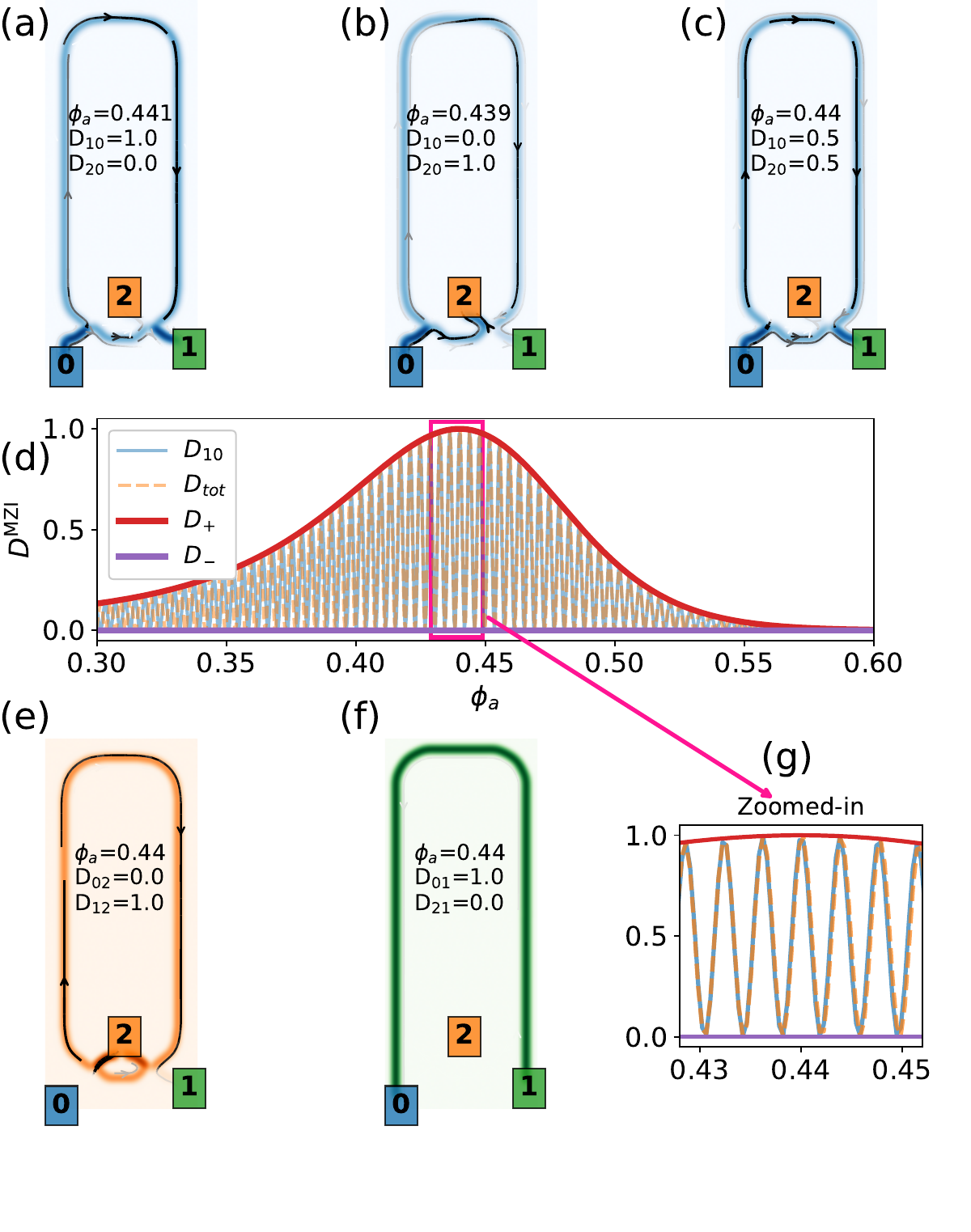}
\caption{
\label{fig:MZI_DC_characteristics}
DC characterization of the MZI model. 
(a),(b) and (c): snapshot of the current density injected from contact $0$ at slightly different magnetic fields. 
The different panels correspond to perfect transmission to contact $1$ ($\phi_a=0.441$, panel a), perfect transmission to contact $2$ ($\phi_a=0.439$, panel b) and half-half
($\phi_a=0.44$, panel c).
(d) Transmission $D_{10}$ versus $\phi_a$. The blue line (simulation) matches with the
theory Eq.\eqref{eq:dc_mzi} (dashed orange). The two envelops $D_+$ (red) and 
$D_-$ (purple) correspond to Eq.\eqref{eq:envelop}.
(e) and (f) snapshot of the current density injected from contact $2$ (panel e) 
and contact $1$ (panel f).
(g) Zoom of panel d. At this scale the theory Eq.\eqref{eq:dc_mzi} is undistinguishable from the numerical data.}
\end{figure}

We are now ready to put all the pieces together and perform the simulations of the full MZI device. First, we check the validity of Eq.\eqref{eq:dc_mzi} in the DC regime. In the numerics (as in actual experiments), we use a uniform magnetic field $\phi_a$ which enters into Eq.\eqref{eq:dc_mzi} in two vastly different scales: the flux $\Phi = \phi_a N_{\text{hole}}$ varies very rapidly with $\phi_a$. Here $N_{\text{hole}} \approx 10,000$
is the number of sites circonvoluted by the two interfering paths \emph{including} the
hole in the center of the sample. A variation of $\phi_a$ of $\pm 0.01$ is sufficient to change the interference from fully destructive to fully constructive as shown in Fig. \ref{fig:MZI_DC_characteristics}
The second role of the magnetic field, at a much larger scale, is to set the system in the quantum Hall
effect and hence define the width of the edge states. It follows that e.g. $D_A(\phi_a)$ and 
$D_B(\phi_a)$ have a slow variation with $\phi_a$ that sets the envelop of the interference pattern between $D_-(\phi_a)$ and $D_+(\phi_a)$ defined as,
\begin{equation}
\label{eq:envelop}
\begin{split}
D_\pm(\phi_a) = D_A D_B + (1-D_A)(1-D_B) \\
\pm 2\sqrt{D_A D_B (1-D_A)(1-D_B)}
\end{split}
\end{equation}
Since the QPC can be simulated separately from the full device, we can compute 
$D_\pm(\phi_a)$ and check that the above formula indeed correspond to the envelop of the interference pattern.

Fig. \ref{fig:MZI_DC_characteristics}(d) shows  the MZI DC transmission $D^\text{MZI}$ versus $\phi_a$. At this scale the interference oscillate very fast and are almost blurred but we can check that the curves
 $D_\pm(\phi_a)$, calculated separately, do match the envelop  of the full calculation. The zoom shown in Fig. \ref{fig:MZI_DC_characteristics}(g) also shows $D_\text{tot}$ from Eq.\eqref{eq:dc_mzi} 
which matches the numerics precisely. However, in Fig. \ref{fig:MZI_DC_characteristics}(d), the value of $N_{\text{hole}}$
has been fitted to match the period of the oscillations of $D^\text{MZI}_{10}$. The obtained value of $N_{\text{hole}}\approx 10000$ is larger than the one obtained by simply looking at the geometry of the sample $N_{\text{hole}}\approx 2500$. This discrepancy is due to our value of $\phi_a\approx 0.44$ being too large to be fully in the continuum limit so that Peierls substitution (strictly speaking valid for $\phi_a\ll 1$) introduces some deviations. We have checked explicitly by running simulations on bigger systems (feasible in DC but not for time-dependent simulations) that this discrepancy disappears as soon as the number of sites (at fixed physical system size) is made four times larger. Fig. \ref{fig:MZI_DC_characteristics}(a),(b),(c) and (e),(f) shows snapshots of the current density, injected from the different contacts (blue for contact 0, green for contact 1 and orange for contact 2) at various values of the magnetic field corresponding to constructive, destructive and mixed interferences (blue snapshots). The entire picture is fully consistent with our understanding of this device working as an electronic Mach-Zehnder interferometer.

\begin{figure}[H]
\includegraphics[width=\linewidth]{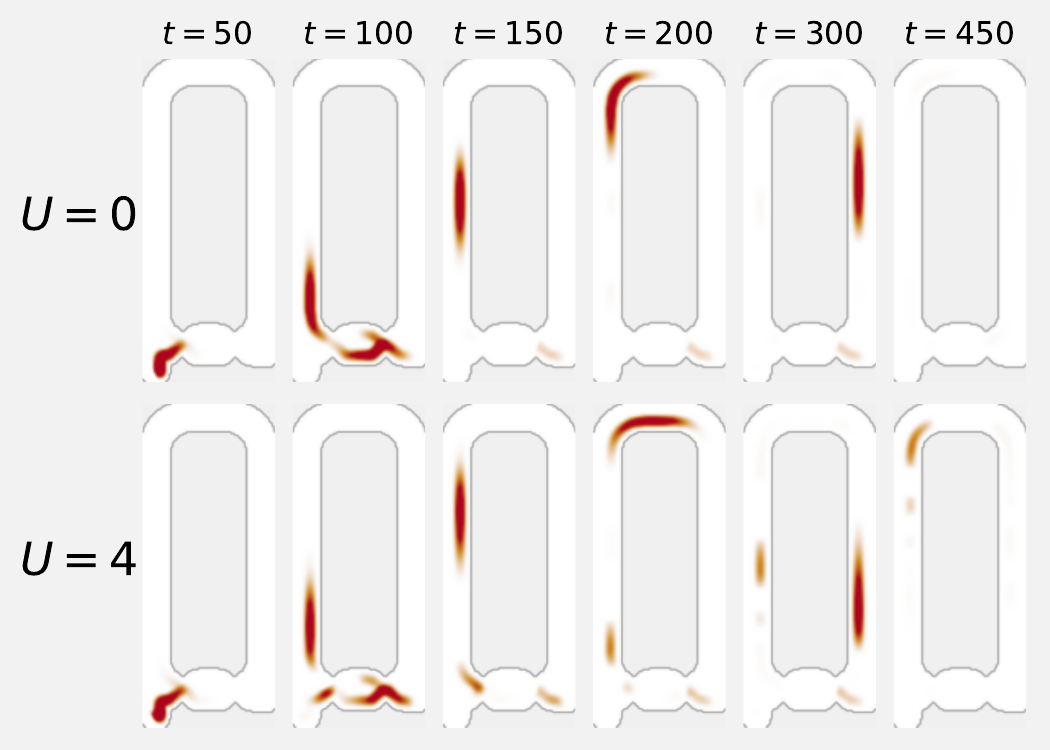}
\caption{\label{fig:MZI_pulse} 
Colormap of the electronic density $\rho(x,y,t)$ within the MZI geometry after injection of a pulse at contact $0$. Upper panels: $U=0$, lower panels: $U=4$.} 
\end{figure}

\begin{figure}[H]
\includegraphics[width=\linewidth]{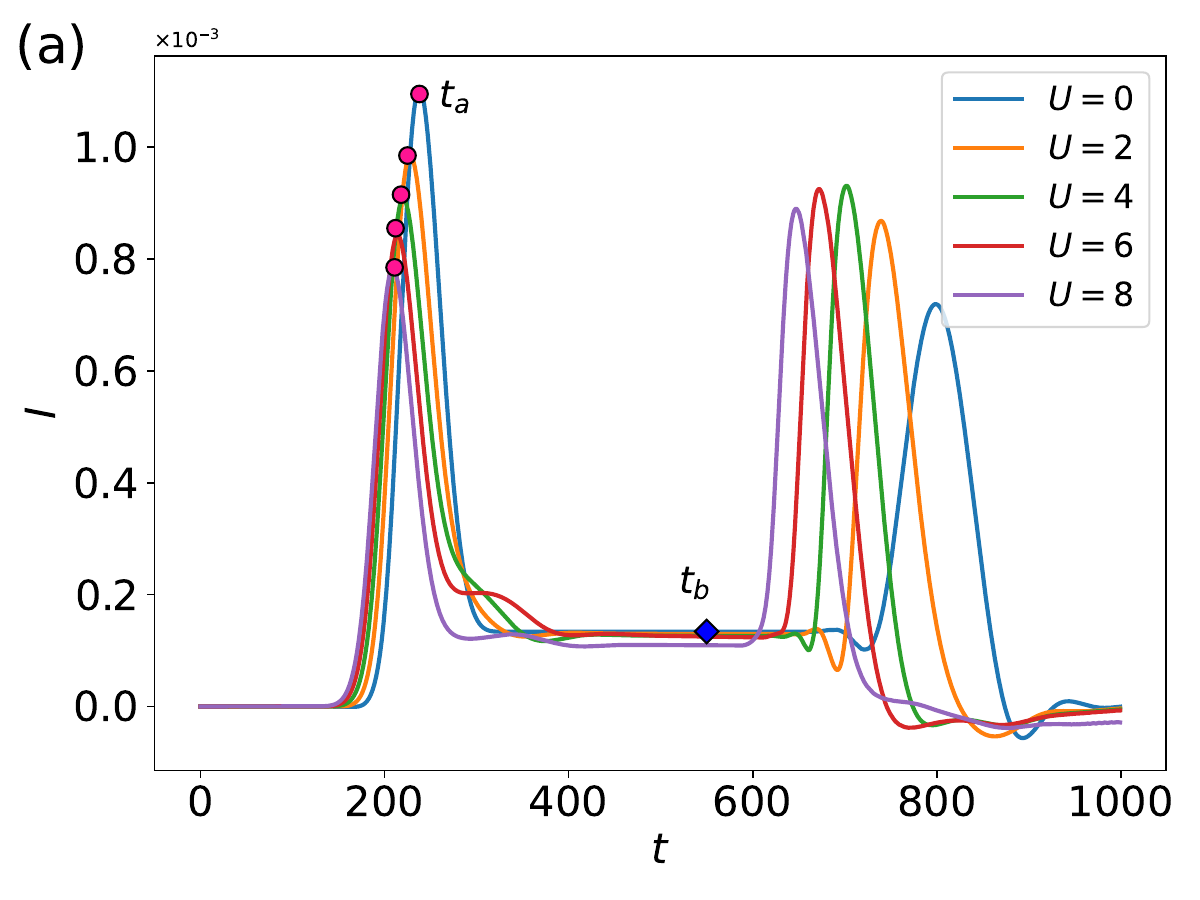} \subfloat{\includegraphics[width=0.5\linewidth]{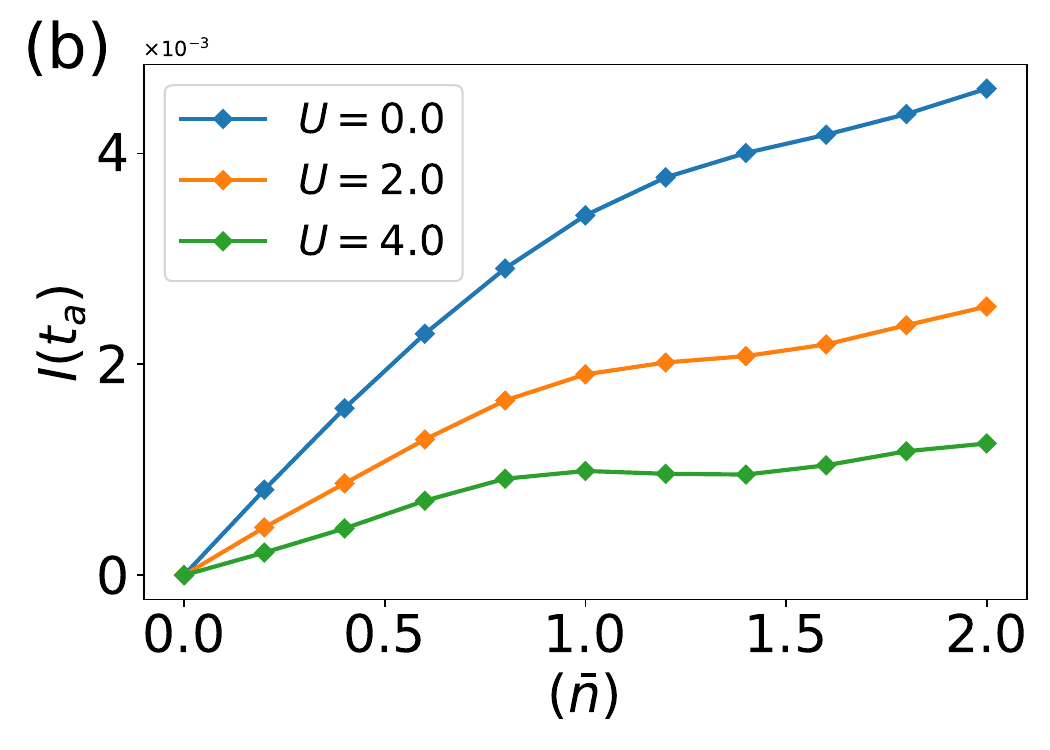}}\subfloat{\includegraphics[width=0.5\linewidth]{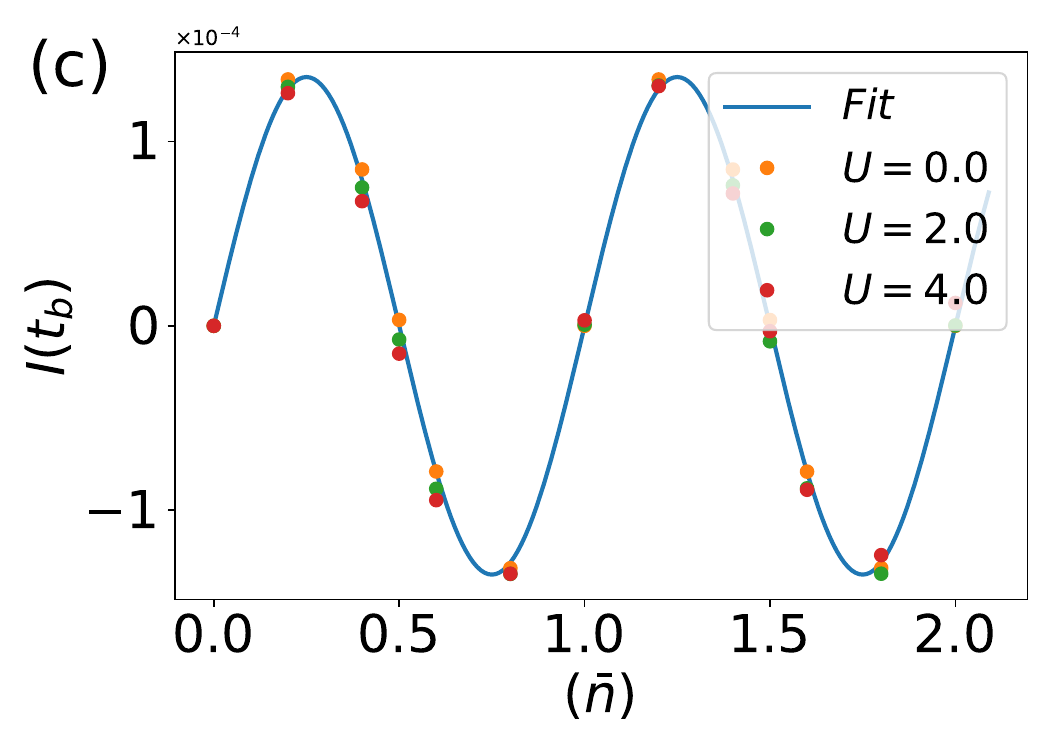}}
\caption{\label{fig:MZI_AC_characteristics}
Current at contact $1$ after pulse injection at contact $0$ in the MZI geometry. 
(a) Current $I(t)$ at contact $1$ versus time for different values of the interaction parameter $U$. The times $t_a$ and $t_b$ correspond respectively to the position of the first peak (arrival of the pulse through the lower arm) and the transient plateau.
(b) Amplitude of the current $I(t_{a})$ versus number of injected charges $\bar n$
(varied by changing the amplitude of the pulse $V_P$ at fixed pulse duration $\tau_P=36/\gamma$). Different curves corresponds to different interaction
strength $U$.
(c) Same as panel b for $I(t_{b})$. The solid line corresponds to a fit
$I = 10^{-4} \sin\left(2\pi\bar{n}\right)$.}
\end{figure}

We now turn to the actual pulse propagation through the device. Snapshots of the excess density at different times are shown in Fig. \ref{fig:MZI_pulse} without (panel a) and with (panel b, $U=4$)
interaction. We observe the propagation of the pulse in the lower and upper arms. With interaction, the situation is a little fuzzy due to the interacting effects at the QPC studied in the preceding section but the global picture is the same with a long transient regime where the pulse has already gone through the lower arms but is still traveling
through the upper one. At a quantitative level, the current passing through contact 1 is shown in Fig. \ref{fig:MZI_AC_characteristics}(a). This curve has three distinctive features already present without interaction: two peaks corresponding to the arrival of the pulse through respectively the lower and upper arm; a long non-zero plateau in the transient regime between these two peaks. The dynamical control of interference pattern effect, of prime interest here, is contained in this plateau. As one increases the value of the peak potential of the pulse $V_P$, thereby changing the total number of electrons $\bar n$ carried in the pulse, the value of this plateau \emph{oscillates} with $\bar n$, as shown in Fig. \ref{fig:MZI_AC_characteristics}(c) while the value of the peak is essentially monotonous (Fig. \ref{fig:MZI_AC_characteristics}(b)).

We observe that the effect of switching on the interaction is two-folds: first, the position of the peaks is translated to the left. This is to be expected from the renormalization of the plasmon velocity discussed in section \ref{sec:pulse_vel}. Second, the shape of these peaks is modified, 
which is consistent with what we found at the QPC level. However, the most important feature of Fig. \ref{fig:MZI_AC_characteristics} is an \emph{absence} of effect: the value of the transient plateau is unaffected by the interaction. This is relevant because this plateau corresponds to interesting new physics that one would wish to be robust to the presence of perturbations.

\begin{figure}[H]
\includegraphics[width=\linewidth, height=0.75\linewidth]{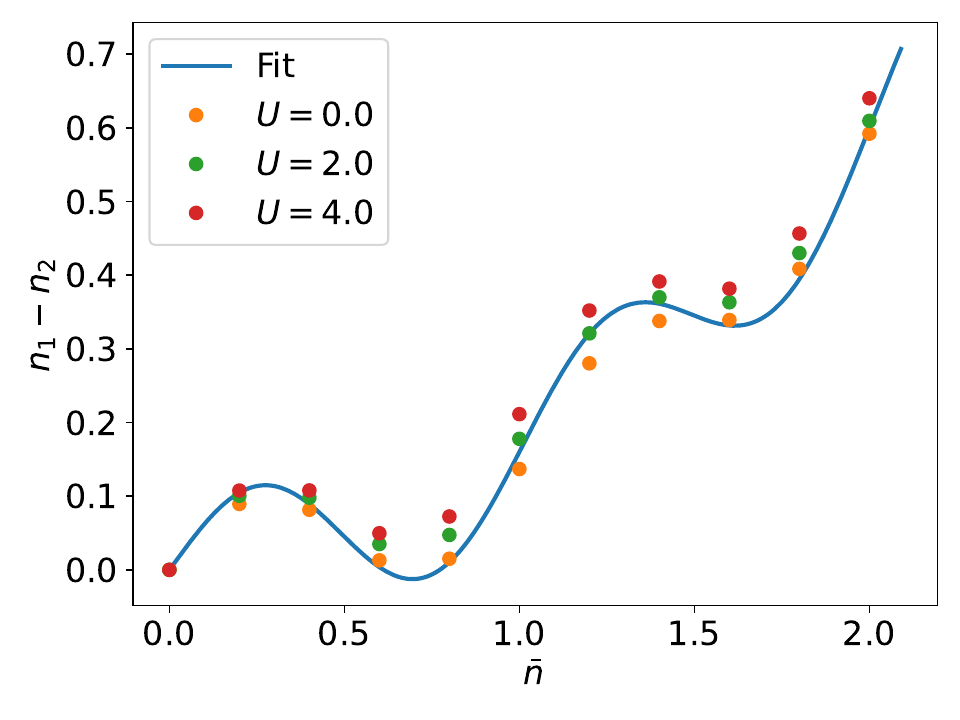}
\includegraphics[width=\linewidth, height=0.75\linewidth]{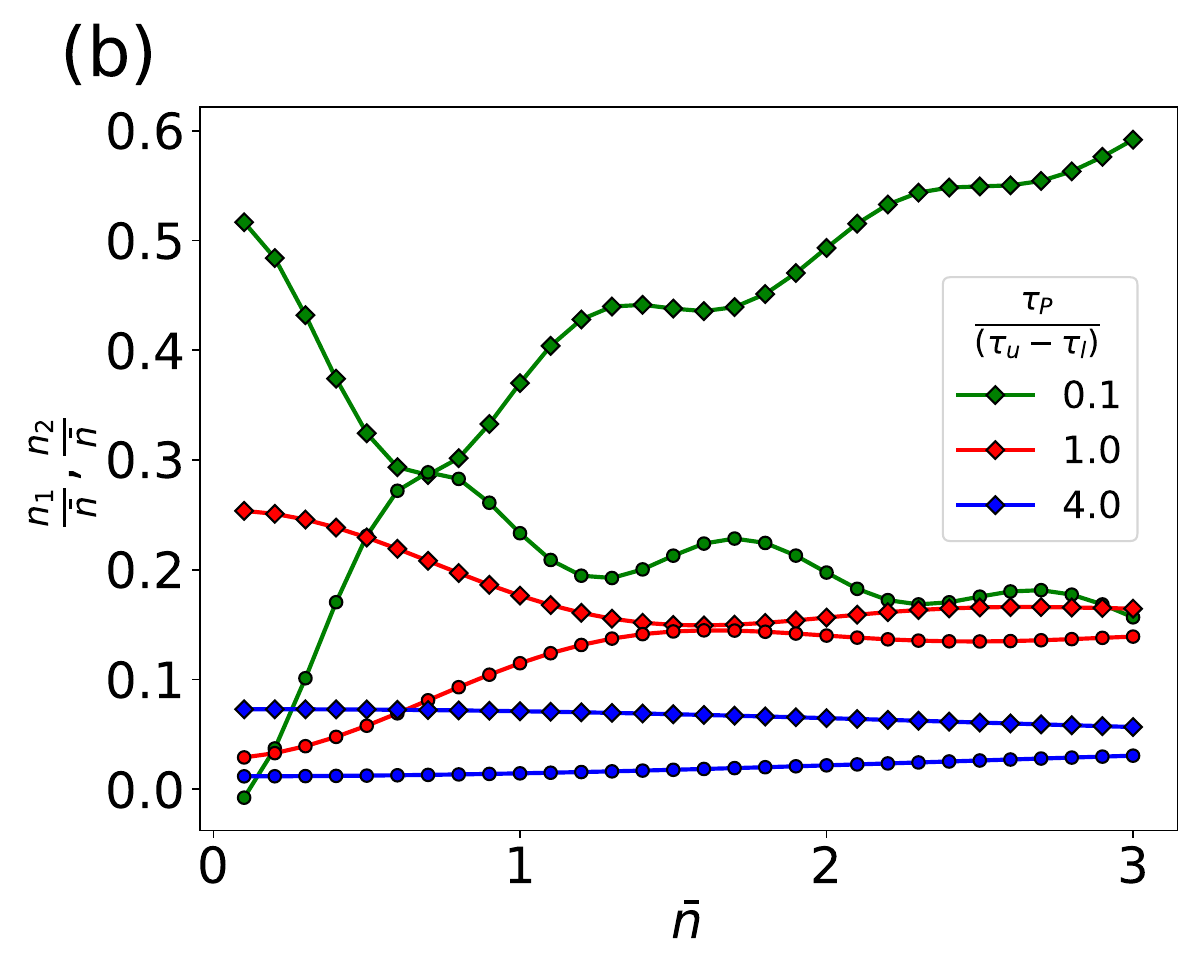}
\caption{\label{fig:MZI_n1_n2_vs_nbar}
Transmitted charge versus injected charge for the full MZI. 
(a) Difference $\left(n_{1}-n_{2}\right)$ between the charge collected at contact $1$
and $2$ versus injected charge $\bar n$ for different values of $U$.
The solid line correspond to a fit 
 $n_{1}-n_{2}=0.02\bar{n}+0.14\bar{n}^{2}+0.10\sin\left(2\pi\bar{n}\right)$
(b) Effective transmission $n_1/\bar n$ (diamonds) and $n_2/\bar n$ (circles) versus $\bar n$ for three different pulse durations and without interaction.
The pulse duration is measured in terms of the size of the transient plateau i.e. the difference of propagation time between the lower arm $\tau_l$ and the upper arm $\tau_u$.
$\tau_P=30/\gamma$ (green),  $\tau_P=337/\gamma$ (red) and  $\tau_P=1348/\gamma$ (blue).}
\end{figure}

Fig. \ref{fig:MZI_n1_n2_vs_nbar} gathers the values of $n_1$ and $n_2$ versus $\bar n$, i.e. the actual observables that would be measured in an experiment. Indeed, in a typical experimental setup, one does not send a single pulse but a string of pulses separated by a period $T$ which has to be chosen large enough for the individual pulses to be considered as independent (in practice
$T$ can be as large as $1$ ns). In this setup, the DC current measured at contact 1 is $en_{1}/T$. Since DC currents are significantly easier to measure than AC ones, especially close to 1 THz, this  is likely to be the first signature of dynamical interference pattern that will be measured in practice. We find that the typical oscillations of $n_1-n_2$ that are present without interaction are barely affected by interaction (Fig. \ref{fig:MZI_n1_n2_vs_nbar}(a)). Note that  these oscillating effects suppose that the pulse duration is short compared to the difference of time of flight $\tau_ u-\tau_l$ between the upper and lower arm. As one increases the 
pulse duration $\tau_P$ the effect disappear (Fig. \ref{fig:MZI_n1_n2_vs_nbar}(b)).

\section{Conclusions}
In this work, we have simulated the propagation of a voltage pulse through an
electronic MZI of a realistic model taking into account the electron-electron
interactions at the time-dependent mean field level. We have observed and characterized the
renormalization of the plasmon velocity (extending previous works to the quantum Hall regime), identified the quantum point contacts as sources of deviations from the non-interacting limit and found that the dynamical control of interference pattern effect is
robust to the presence of interactions. Overall, this study gives a rather optimistic
picture that the most salient features predicted from the non-interacting theory might actually be present in real devices. In fact preliminary experimental results \cite{BauerlePrivateComm} seem to indicate that it is the case.

On the theory side, much work remains to be done to arrive at the level where the simulations can have genuine predictive power. One aspect is to replace the contact interaction with the solution of the Poisson equation (in real time). This extension would make the potential seen by the electrons calculated instead of postulated which would make contact with experiments much easier. Also, effects such as edge reconstruction \cite{Armagnat2019, Chklovskii1992}  would be accounted for automatically. Such a step will be computationally expensive but should remain feasible due to the fact that the electric potential varies at a very slow time scale compared to the wavefunctions. An iterative solver, that would benefit from the fact that the solution at a slightly earlier time can be used as a starting point, might be the method of choice. A second aspect is to introduce decoherence. In DC, it is known that decoherence is controlled by one particle decaying into two particles and one hole, i.e. by a second order contribution to the self energy in the strength of electron-electron interaction. Such a contribution could be calculated using e.g. tensor network approaches in the spirit of \cite{NezFernndez2022}. Studying what happens to decoherence in the THz regime will be of the utmost interest and is crucial for future applications. A third extension would be the study of abelian anyons that could be done within a framework close to the present one using a composite fermion approach
\cite{Halperin1993}. This would allow to make concrete predictions in e.g. the 1/3 fraction
or even in more exotic fractions. Last, a systematic comparison between the simulations and the experiments, based on large data sets, should be performed to validate the overall approach (in the spirit of \cite{Chatzikyriakou2022}).

\section{Acknowledgments}

P.K., T.K. and X.W. acknowledge funding from the European Union’s Horizon 2020 research and innovation program under Grant agreement No. 862683 (UltraFastNano), from the French ANR DADDI and T-KONDO and from State aid managed by the Agence Nationale de la Recherche under the France 2030 program, reference ANR-22-PETQ-0012 (EQUBITFLY). T.K. greatly appreciates the hospitality of C. Bauerle in his group. 

\bibliography{mean_field}

@misc{Tkwant2021,
  title        = {Tkwant: A Python package for time-dependent quantum transport},
  year         = {2021},
  howpublished = {\url{https://tkwant.kwant-project.org/}}
}

@article{Kloss2025,
   author = {Thomas Kloss and Xavier Waintal},
   doi = {10.1103/PhysRevB.111.235411},
   issn = {2469-9950},
   issue = {23},
   journal = {Physical Review B},
   month = {6},
   pages = {235411},
   title = {Propagation of ultrashort voltage pulses through a small quantum dot},
   volume = {111},
   year = {2025}
}

@article{Saha2025,
   author = {Mrinmoyee Saha and Luca Horray and Pedro Portugal and Christian Flindt},
   doi = {10.1103/36rq-15v7},
   issn = {2469-9950},
   issue = {8},
   journal = {Physical Review B},
   month = {8},
   pages = {L081408},
   title = {Negative currents in Fabry-Pérot cavities are caused by interfering paths},
   volume = {112},
   year = {2025}
}

@article{Portugal2024,
   author = {Pedro Portugal and Fredrik Brange and Christian Flindt},
   doi = {10.1103/PhysRevLett.132.256301},
   issn = {0031-9007},
   issue = {25},
   journal = {Physical Review Letters},
   month = {6},
   pages = {256301},
   title = {Heat Pulses in Electron Quantum Optics},
   volume = {132},
   year = {2024}
}

@article{Chatzikyriakou2022,
   author = {Eleni Chatzikyriakou and Junliang Wang and Lucas Mazzella and Antonio Lacerda-Santos and Maria Cecilia da Silva Figueira and Alex Trellakis and Stefan Birner and Thomas Grange and Christopher Bäuerle and Xavier Waintal},
   doi = {10.1103/PhysRevResearch.4.043163},
   issn = {2643-1564},
   issue = {4},
   journal = {Physical Review Research},
   month = {12},
   pages = {043163},
   title = {Unveiling the charge distribution of a GaAs-based nanoelectronic device: A large experimental dataset approach},
   volume = {4},
   year = {2022}
}

@article{NezFernndez2022,
   author = {Yuriel Núñez Fernández and Matthieu Jeannin and Philipp T. Dumitrescu and Thomas Kloss and Jason Kaye and Olivier Parcollet and Xavier Waintal},
   doi = {10.1103/PhysRevX.12.041018},
   issn = {2160-3308},
   issue = {4},
   journal = {Physical Review X},
   month = {11},
   pages = {041018},
   title = {Learning Feynman Diagrams with Tensor Trains},
   volume = {12},
   year = {2022}
}

@article{Kloss2021,
   author = {Thomas Kloss and Joseph Weston and Benoit Gaury and Benoit Rossignol and Christoph Groth and Xavier Waintal},
   doi = {10.1088/1367-2630/abddf7},
   issn = {1367-2630},
   issue = {2},
   journal = {New Journal of Physics},
   month = {2},
   pages = {023025},
   title = {Tkwant: a software package for time-dependent quantum transport},
   volume = {23},
   year = {2021}
}

@article{Armagnat2019,
   author = {Pacome Armagnat and A. Lacerda-Santos and Benoit Rossignol and Christoph Groth and Xavier Waintal},
   doi = {10.21468/SciPostPhys.7.3.031},
   issn = {2542-4653},
   issue = {3},
   journal = {SciPost Physics},
   month = {9},
   pages = {031},
   title = {The self-consistent quantum-electrostatic problem in strongly non-linear regime},
   volume = {7},
   year = {2019}
}

@article{Roussely2018,
   author = {Gregoire Roussely and Everton Arrighi and Giorgos Georgiou and Shintaro Takada and Martin Schalk and Matias Urdampilleta and Arne Ludwig and Andreas D. Wieck and Pacome Armagnat and Thomas Kloss and Xavier Waintal and Tristan Meunier and Christopher Bäuerle},
   doi = {10.1038/s41467-018-05203-7},
   issn = {2041-1723},
   issue = {1},
   journal = {Nature Communications},
   month = {7},
   pages = {2811},
   title = {Unveiling the bosonic nature of an ultrashort few-electron pulse},
   volume = {9},
   year = {2018}
}

@article{Bauerle2018,
   author = {Christopher Bäuerle and D Christian Glattli and Tristan Meunier and Fabien Portier and Patrice Roche and Preden Roulleau and Shintaro Takada and Xavier Waintal},
   doi = {10.1088/1361-6633/aaa98a},
   issn = {0034-4885},
   issue = {5},
   journal = {Reports on Progress in Physics},
   month = {5},
   pages = {056503},
   title = {Coherent control of single electrons: a review of current progress},
   volume = {81},
   year = {2018}
}

@article{Kloss2018,
   author = {Thomas Kloss and Joseph Weston and Xavier Waintal},
   doi = {10.1103/PhysRevB.97.165134},
   issn = {2469-9950},
   issue = {16},
   journal = {Physical Review B},
   month = {4},
   pages = {165134},
   title = {Transient and Sharvin resistances of Luttinger liquids},
   volume = {97},
   year = {2018}
}

@article{Gaury2015,
   author = {Benoit Gaury and Joseph Weston and Xavier Waintal},
   doi = {10.1038/ncomms7524},
   issn = {2041-1723},
   issue = {1},
   journal = {Nature Communications},
   month = {3},
   pages = {6524},
   title = {The a.c. Josephson effect without superconductivity},
   volume = {6},
   year = {2015}
}

@article{Gaury2014b,
   author = {Benoit Gaury and Xavier Waintal},
   doi = {10.1038/ncomms4844},
   issn = {2041-1723},
   issue = {1},
   journal = {Nature Communications},
   month = {5},
   pages = {3844},
   title = {Dynamical control of interference using voltage pulses in the quantum regime},
   volume = {5},
   year = {2014}
}

@article{Gaury2014a,
   author = {Benoit Gaury and Joseph Weston and Xavier Waintal},
   doi = {10.1103/PhysRevB.90.161305},
   issn = {1098-0121},
   issue = {16},
   journal = {Physical Review B},
   month = {10},
   pages = {161305},
   title = {Stopping electrons with radio-frequency pulses in the quantum Hall regime},
   volume = {90},
   year = {2014}
}

@article{Groth2014,
   author = {Christoph W Groth and Michael Wimmer and Anton R Akhmerov and Xavier Waintal},
   doi = {10.1088/1367-2630/16/6/063065},
   issn = {1367-2630},
   issue = {6},
   journal = {New Journal of Physics},
   month = {6},
   pages = {063065},
   title = {Kwant: a software package for quantum transport},
   volume = {16},
   year = {2014}
}

@article{Roulleau2008,
   author = {Preden Roulleau and F. Portier and P. Roche and A. Cavanna and G. Faini and U. Gennser and D. Mailly},
   doi = {10.1103/PhysRevLett.100.126802},
   issn = {0031-9007},
   issue = {12},
   journal = {Physical Review Letters},
   month = {3},
   pages = {126802},
   title = {Direct Measurement of the Coherence Length of Edge States in the Integer Quantum Hall Regime},
   volume = {100},
   year = {2008}
}

@article{Buttiker1995,
   author = {M. Büttiker},
   doi = {10.1007/BF02741461},
   issn = {1826-9877},
   issue = {5-6},
   journal = {Il Nuovo Cimento B},
   month = {5},
   pages = {509-522},
   title = {Time-dependent current partition in mesoscopic conductors},
   volume = {110},
   year = {1995}
}

@article{Buttiker1993,
   author = {M Buttiker},
   doi = {10.1088/0953-8984/5/50/017},
   issn = {0953-8984},
   issue = {50},
   journal = {Journal of Physics: Condensed Matter},
   month = {12},
   pages = {9361-9378},
   title = {Capacitance, admittance, and rectification properties of small conductors},
   volume = {5},
   year = {1993}
}

@article{Heiblum2003,
  author  = {Ji, Y. and Chung, Y. and Sprinzak, D. and Heiblum, M. and Mahalu, D. and Shtrikman, H.},
  title   = {An electronic Mach--Zehnder interferometer},
  journal = {Nature},
  volume  = {422},
  pages   = {415--418},
  year    = {2003},
  doi     = {10.1038/nature01503}
}

@article{Chklovskii1992,
  title = {Electrostatics of edge channels},
  author = {Chklovskii, D. B. and Shklovskii, B. I. and Glazman, L. I.},
  journal = {Phys. Rev. B},
  volume = {46},
  issue = {7},
  pages = {4026--4034},
  numpages = {0},
  year = {1992},
  month = {Aug},
  publisher = {American Physical Society},
  doi = {10.1103/PhysRevB.46.4026},
  url = {https://link.aps.org/doi/10.1103/PhysRevB.46.4026}
}

@article{Halperin1993,
  title = {Theory of the half-filled Landau level},
  author = {Halperin, B. I. and Lee, Patrick A. and Read, Nicholas},
  journal = {Phys. Rev. B},
  volume = {47},
  issue = {12},
  pages = {7312--7343},
  numpages = {0},
  year = {1993},
  month = {Mar},
  publisher = {American Physical Society},
  doi = {10.1103/PhysRevB.47.7312},
  url = {https://link.aps.org/doi/10.1103/PhysRevB.47.7312}
}

@misc{Xavier2024,
  title        = {Computational quantum transport},
  author       = {Waintal, Xavier and Wimmer, Michael and Akhmerov, Anton and Groth, Christoph and Nikoli{\'c}, Branislav K. and Istas, Mathieu and Rosdahl, T{\'o}mas {\"O}rn and Varjas, Daniel},
  year         = {2024},
  month        = {July},
  eprint       = {2407.16257},
  archivePrefix = {arXiv},
  primaryClass = {cond-mat.mes-hall},
  note         = {arXiv preprint arXiv:2407.16257}
}

@article{Micolich2011,
  author       = {A. P. Micolich},
  title        = {What lurks below the last plateau: Experimental studies of the 0.7 × 2$e^{2}$/h conductance anomaly in one-dimensional systems},
  journal      = {Journal of Physics: Condensed Matter},
  year         = {2011},
  volume       = {23},
  number       = {44},
  pages        = {443201},
  month        = nov,
  doi          = {10.1088/0953-8984/23/44/443201},
  issn         = {0953-8984}
}

@book{Negele2018,
   author = {John W. Negele and Henri Orland},
   doi = {10.1201/9780429497926},
   isbn = {9780429497926},
   month = {3},
   publisher = {CRC Press},
   title = {Quantum Many-Particle Systems},
   year = {2018}
}

@article{Nakamura2020,
   author = {J. Nakamura and S. Liang and G. C. Gardner and M. J. Manfra},
   doi = {10.1038/s41567-020-1019-1},
   issn = {1745-2473},
   issue = {9},
   journal = {Nature Physics},
   month = {9},
   pages = {931-936},
   title = {Direct observation of anyonic braiding statistics},
   volume = {16},
   year = {2020}
}

@article{Nakamura2022,
   author = {J. Nakamura and S. Liang and G. C. Gardner and M. J. Manfra},
   doi = {10.1038/s41467-022-27958-w},
   issn = {2041-1723},
   issue = {1},
   journal = {Nature Communications},
   month = {1},
   pages = {344},
   title = {Impact of bulk-edge coupling on observation of anyonic braiding statistics in quantum Hall interferometers},
   volume = {13},
   year = {2022}
}

@misc{BauerlePrivateComm,
  author       = {Chris Bäuerle},
  title        = {Private communication},
  year         = {2025}
}

@article{Levitov1996,
  title     = {Electron counting statistics and coherent states of electric current},
  author    = {Levitov, L. S. and Lee, H.-W. and Lesovik, G. B.},
  journal   = {Journal of Mathematical Physics},
  volume    = {37},
  number    = {10},
  pages     = {4845--4866},
  year      = {1996},
  publisher = {AIP Publishing},
  doi       = {10.1063/1.531672},
  url       = {https://doi.org/10.1063/1.531672}
}

@article{Dubois2013,
  title     = {Minimal-excitation states for electron quantum optics using levitons},
  author    = {Dubois, J. and Jullien, T. and Portier, F. and Roche, P. and Cavanna, A. and Jin, Y. and Wegscheider, W. and Roulleau, P. and Glattli, D. C.},
  journal   = {Nature},
  year      = {2013},
  month     = {October},
  volume    = {502},
  number    = {7473},
  pages     = {659--663},
  doi       = {10.1038/nature12713},
  url       = {https://doi.org/10.1038/nature12713},
  issn      = {1476-4687}
}

@article{Levitov2006,
  title = {Minimal Excitation States of Electrons in One-Dimensional Wires},
  author = {Keeling, J. and Klich, I. and Levitov, L. S.},
  journal = {Phys. Rev. Lett.},
  volume = {97},
  issue = {11},
  pages = {116403},
  numpages = {4},
  year = {2006},
  month = {Sep},
  publisher = {American Physical Society},
  doi = {10.1103/PhysRevLett.97.116403},
  url = {https://link.aps.org/doi/10.1103/PhysRevLett.97.116403}
}

@article{Levitov1997,
  title = {Coherent states of alternating current},
  author = {Ivanov, D. A. and Lee, H. W. and Levitov, L. S.},
  journal = {Phys. Rev. B},
  volume = {56},
  issue = {11},
  pages = {6839--6850},
  numpages = {0},
  year = {1997},
  month = {Sep},
  publisher = {American Physical Society},
  doi = {10.1103/PhysRevB.56.6839},
  url = {https://link.aps.org/doi/10.1103/PhysRevB.56.6839}
}

\end{document}